\documentclass[twocolumn]{aastex63}

\usepackage{mathrsfs}

\accepted{to ApJ on 16 December 2022}

\usepackage[morefloats=10]{morefloats}

\shorttitle{AGN feedback driven quenching}
\shortauthors{Bluck et al.}

\begin{document}

\title{The fundamental signature of star formation quenching from AGN feedback: \\ A critical dependence of quiescence on supermassive black hole mass not accretion rate}


\correspondingauthor{Asa F. L. Bluck}
\email{abluck@fiu.edu}

\author[0000-0001-6395-4504]{Asa F. L. Bluck}
\affiliation{Department of Physics, Florida International University,11200 SW 8th Street, Miami, 33199, Florida, USA}

\author[0000-0003-1661-2338]{Joanna M. Piotrowska} 
\affiliation{Kavli Institute for Cosmology, University of Cambridge, Madingley Road, Cambridge, CB3 0HA, UK}
\affiliation{Cavendish Laboratory - Astrophysics Group, University of Cambridge, 19 JJ Thomson Avenue, Cambridge, CB3 0HE, UK}

\author[0000-0002-4985-3819]{Roberto Maiolino}
\affiliation{Kavli Institute for Cosmology, University of Cambridge, Madingley Road, Cambridge, CB3 0HA, UK}
\affiliation{Cavendish Laboratory - Astrophysics Group, University of Cambridge, 19 JJ Thomson Avenue, Cambridge, CB3 0HE, UK}
\affiliation{Department of Physics and Astronomy, University College London, Gower Street, London WC1E 6BT, UK}


\begin{abstract}

\noindent We identify the intrinsic dependence of star formation quenching on a variety of galactic and environmental parameters, utilizing a machine learning approach with Random Forest classification. We have previously demonstrated the power of this technique to isolate causality, not mere correlation, in complex astronomical data. First, we analyze three cosmological hydrodynamical simulations (Eagle, Illustris, and IllustrisTNG), selecting snapshots spanning the bulk of cosmic history from comic noon ($z \sim 2$) to the present epoch, with stellar masses in the range $9 < \log(M_*/M_{\odot}) < 12$. In the simulations, black hole mass is unanimously found to be the most predictive parameter of central galaxy quenching at all epochs. Perhaps surprisingly, black hole accretion rate (and hence the bolometric luminosity of active galactic nuclei, AGN) is found to be of little predictive power over quenching. This theoretical result is important for observational studies of galaxy quenching as it cautions against using the current AGN state of a galaxy as a useful proxy for the cumulative impact of AGN feedback on a galactic system. The latter is traced by black hole mass {\it not} AGN luminosity. Additionally, we explore a sub-set of ‘observable' parameters, which can be readily measured in extant wide-field galaxy surveys targeting $z = 0 - 2$, at $9 < \log(M_*/M_{\odot}) < 12$. All three simulations predict that in lieu of black hole mass, the stellar gravitational potential will outperform the other parameters in predicting quenching. We confirm this theoretical prediction observationally in the SDSS (at low redshifts) and in CANDELS (at intermediate and high redshifts). 

\end{abstract}

\keywords{Galaxies: formation, evolution, star formation, quenching, feedback}


\section{Introduction}

\noindent Galaxies are observed to display profound bimodality in key diagnostic diagrams, most notably in color - magnitude (e.g., Strateva et al. 2001). Perhaps more physically, the star forming main sequence relation between star formation rate (SFR) and stellar mass ($M_*$) exhibits a tight star forming ridge line, with quiescent galaxies significantly offset to lower star formation rates at a fixed stellar mass (e.g., Brinchmann et al. 2004). Consequently, it is often stated that galaxies exhibit two fundamental types: i) actively star forming, and ii) quiescent (or ‘quenched') systems. Quenched systems tend to have redder optical colors, older stellar populations, more elliptical morphologies, more pressure supported kinematics, higher masses, and reside in denser cosmic environments compared to their actively star forming counterparts (e.g., Baldry et al. 2006; Peng et al. 2010, 2012; Wuyts et al. 2011; Bluck et al. 2014, 2016; Brownson et al. 2022). Much of the field of galaxy evolution is explicitly focused on explaining the origin of quiescent galaxies, and hence accounting for the observations of bimoodality in galactic properties.

Theoretically, the existence of quiescent galaxies is translated to a problem of cosmological star formation efficiency. More precisely, one of the main goals of modern simulations of galaxy formation is to reduce the efficiency with which stars form within dark matter haloes (i.e., $ \epsilon_{SF} \equiv M_*/M_{\rm Halo} \leq 0.1\,\Omega_b/\Omega_M$). The reason for this is that simple models of galaxy formation (utilising only gravitation and cooling) predict that the vast majority of baryons should reside in stars by the present epoch (e.g., Cole et al. 2000; Bower et al. 2006, 2008; Henriques et al. 2015, 2019). However, observations place the fraction of baryons in stars an order of magnitude lower (e.g., Fukugita \& Peebles 2004; Shull 2012). In the past decade or so, feedback from active galactic nuclei (AGN) has become the favored mechanism in modern hydrodynamical simulations to quench star formation, and hence reproduce the observed galaxy bimodality (see, e.g., Sijacki et al. 2007; Vogelsberger et al. 2014a,b; Schaye et al. 2015; Weinberger et al. 2018; Zinger et al. 2020). There is abundant evidence that AGN produce more than enough energy to quench star formation (e.g., Silk \& Rees 1998; Bluck et al. 2011; Maiolino et al. 2012; Fabian 2012). Yet, direct observational evidence for AGN feedback causing quenching within galaxies remains sparse, and hotly debated (although see Hlavacek-Larrondo et al. 2012, 2017, 2018 for perhaps the strongest direct evidence to date, primarily within galaxy clusters and very massive groups). Thus, it remains unclear whether the crucial ingredient in cosmological models to quench galaxies is viable for the vast majority of systems, or not.

Classification is the branch of data science that focuses on understanding the fundamental differences between types of objects, with the ultimate purpose of accurate segregation between them. As such, classification is at the heart of the scientific study of galaxy populations: If we can learn to accurately classify star forming and quenched galaxies on the basis of their physical properties, we can establish the underlying physics of galaxy quenching. For example, we have demonstrated the power of machine learning classification to `reverse engineer' cosmological simulations, revealing the input physics and its observational consequences (see Bluck et al. 2022; Piotrowska et al. 2022). But the real power of machine learning in this application is apparent when one compares the results of classification between simulations and observational data. If there is strong agreement, the models can be used as an explanatory tool for the observations. Alternatively, if there is major disagreement, the observations can be used to improve the next generation of models by ruling out certain prescriptions or processes. 

The goal of this paper is to leverage the power of classification to determine what is intrinsically important for central galaxy quenching in both simulations and observations. To this end, we employ a Random Forest classifier, which enables the identification of causality through carefully controlling for nuisance variables (see Bluck et al. 2022 for a detailed description of this technique including multiple tests).  

Utilizing Random Forest classification, in Piotrowska et al. (2022) we discovered that the most effective parameter for separating star forming and quenched central galaxies in the local Universe is black hole mass. Interestingly, this is a unanimous prediction from three contemporary cosmological simulations (Eagle, Illustris, and IllustrisTNG), which use very different AGN feedback prescriptions. As such, this prediction is not strongly model dependent and hence can be seen as a ubiquitous consequence of AGN feedback (in almost any mode). We also considered stellar mass, dark matter halo mass, and black hole accretion rate as potential drivers of quenching in the simulations. Yet, all of these parameters are of negligible importance to quenching once black hole mass is made available to the classifier. 

The results from Piotrowska et al. (2022) are highly important for observational studies of AGN feedback because they clearly imply that ‘catching AGN quenching in the act' is not possible in the local Universe. Instead, one must look for the fossil record of historic AGN feedback, encapsulated in the mass of the central black hole, {\it not} the current accretion rate or AGN luminosity. We also established that once black hole mass is controlled for, neither stellar nor halo mass should be constraining of central galaxy quenching. We confirmed this latter prediction observationally through comparison with the SDSS, which effectively rules out virial shocks or supernova feedback as significant mechanisms for quenching central galaxies in the local Universe (see also Bluck et al. 2016, 2020a for further evidence on this). 

In this paper, we expand on the work of Piotrowska et al. (2022) by identifying the key quenching predictions from Eagle, Illustris, and IllustrisTNG at multiple epochs from z = 0 to cosmic noon (where the number density of quiescent galaxies becomes very low). This is essential in order to establish whether there is redshift dependence on the predicted signature of AGN feedback quenching or not. For instance, it could be that at high redshifts a strong dependence of quiescence on the current AGN state of a galaxy becomes apparent, which would have profound implications for how to search for observational evidence of AGN feedback (or lack thereof) in upcoming observational surveys (e.g., with JWST and VLT-MOONS).  Additionally, we perform a preliminary test on the high-$z$ predictions from cosmological simulations, utilizing photometric data from HST-CANDELS at z = 0.5 - 2, and compare this to a novel test of the simulations' predictions at low-$z$ utilizing spectroscopic data from the SDSS.

The paper is structured as follows. In Section 2 we give an overview of the simulations and observational data. In Section 3 we present our method to identify quiescent systems and give a brief overview of our Random Forest classification technique. In Section 4 we present our results and discuss their importance for the field of galaxy evolution. We summarize our contributions in Section 5. Additionally, in the Appendix, we present numerous detailed tests on the Random Forest results, which confirm our main conclusions. Throughout the paper we assume a spatially flat $\Lambda$CDM cosmology, and set h $\equiv H_0 / (100$ km/s/Mpc) = 0.7 consistently for all physical representations of simulated and observational data.


\section{Data Sources}

\subsection{Simulations}

\noindent In this paper, we consider three cosmological hydrodynamical simulations which incorporate AGN feedback to quench central (and more generally, massive) galaxies. Our goal is to identify the testable consequences of AGN feedback driven quenching in these models. Explicitly, we work with publicly available multi-epoch snapshot data from: Eagle\footnote{Eagle Data Access: http://icc.dur.ac.uk/Eagle/} (Schaye et al. 2015; Crain et al. 2015; McAlpine et al. 2016); Illustris\footnote{Illustris Data Access: www.illustris-project.org} (Vogelsberger et al. 2014a,b; Genel et al. 2014; Sijacki et al. 2015; Nelson et al. 2016); and IllustrisTNG\footnote{IllustrisTNG Data Access: www.tng-project.org/} (Marinacci et al. 2018; Naiman et al. 2018; Nelson et al. 2018; Springel et al. 2018; Pillepich et al. 2018; Nelson et al. 2019). Full details on the simulations are provided in the above references, including information on data access. A detailed description of the similarities and differences of these three simulations is provided in Piotrowska et al. (2022). Here we give a review of the most important details for this work (i.e. the black hole growth and AGN feedback mechanisms).

In all simulations we select central galaxies as the most massive systems in each (group) dark matter halo. Isolated galaxies are treated as the centrals of their individual dark matter haloes. We select galaxies to have stellar masses in the range $9 < \log(M_*/M_{\odot}) < 12$, residing in dark matter (group) haloes with $M_{\rm halo} > 10^{11} M_{\odot}$. These cuts mitigate issues with mass and volume resolution in the simulations, whilst still enabling a large sampling of both star forming and quenched systems. Even though the majority of quiescent galaxies in all of the simulations and observations considered in this work have $\log(M_*/M_{\odot}) > 10$, it is vital to select a large sample of both star forming and quiescent classes for random forest classification. All specific parameters are collated as in Piotrowska et al. (2022), but here extracted for multiple snapshots (rather than just at z = 0). The identification of quiescent systems is discussed in Section 3. 

From Illustris and IllustrisTNG (which have the same naming conventions) we take the following parameters from the public snapshot data at z = (0, 0.5, 1, 1.5, 2): From the {\it SubFind} catalog - {\it SubhaloSFR}, {\it SubhaloMassStar}, {\it SubhaloBHMass}, {\it SubhaloBHmdot}, {\it SubhaloHalfmassRad}; and from the {\it FoF} catalog - {\it Group\_M\_Crit200}, {\it Group\_R\_Crit200}. We also extract the co-moving coordinates for each system (galaxy and halo), as well as the {\it FoF} Halo ID and {\it SubFind} sub-halo ID. From Eagle we take the following parameters from the public snapshot data at z = (0, 0.5, 1, 1.49, 2.01): From the {\it SubFind} catalog - {\it StarFormationRate}, {\it MassType\_Star}, {\it BlackHoleMass}, {\it BlackHoleAccretionRate}, {\it HalfMassRad\_Star}; and from the {\it FoF} catalog - {\it Group\_M\_Crit200}, {\it Group\_R\_Crit200}. As with Illustris and IllustrisTNG, we also extract halo and sub-halo coordinates and IDs for each system in both the {\it FoF} and {\it SubFind} catalogs. A publicly available docker is provided with Piotrowska et al. (2022) showing how to extract these parameters from each simulation for the z = 0 snapshots.

\subsubsection{Eagle}

\noindent For Eagle, we utilize the {\small EAGLE-RefL0100N1504} run (Schaye et al. 2015). This run has a box size of $\sim$100 cMpc$^3$ and implements the most detailed feedback mechanisms of the Eagle simulation suite. Eagle is performed utilizing a smoothed particle hydrodynamics (SPH) code, explicitly {\it \small GADGET-3} (Springel 2005). Cosmological parameters are taken from Planck Collaboration I (2014), assuming a spatially flat $\Lambda$CDM cosmology. Black holes are seeded at $M_{BH} = 10^5 M_{\odot}h^{-1}$ in all haloes once they reach $M_{\rm Halo} = 10^{10} M_{\odot}h^{-1}$. Black hole accretion is regulated by Bondi-Hoyle accretion, i.e. $\dot{M}_{BH} \propto M_{BH}^2$ (e.g., Hoyle \& Lyttleton 1936; Bondi \& Hoyle 1944), and is Eddington limited. A single feedback mode is applied, which corresponds approximately to a quasar wind, triggered primarily by cold-mode accretion (see Booth \& Schaye 2009). Energy injection into the surrounding gas particles, in a given time step $\Delta t$, is given explicitly by $\Delta E_{BH} = \epsilon_f \epsilon_r \dot{M}_{BH} c^2 \Delta t$, where $\epsilon_r$ is the radiative efficiency (set equal to 0.1) and $\epsilon_f$ is the fraction of energy which couples to the inter-stellar medium (ISM) producing energetic feedback. Energy is released thermally once $\Delta E_{BH}$ is sufficient to induce a temperature change of $\Delta T = 10^{8.5} K$ for at least one neighboring gas particle. Hence, heating of gas particles near to the black hole is applied stochastically. The large thermal injection, in essentially random directions, is key to overcoming the numerical over-cooling problem in this simulation. The Eagle AGN feedback mechanism is effective at quenching massive galaxies; however, it is less effective at keeping massive galaxies quenched than the other simulations considered here (see, e.g., Piotrowska et al. 2022).

\subsubsection{Illustris}

\noindent For Illustris, we utilize the full {\small ILLUSTRIS-1} run with box size $\sim$100 cMpc$^3$ (Vogelsberger et al. 2014a,b). Illustris is run utilizing the moving-mesh code {\it \small AREPO} (Springel 2010). Cosmological parameters are set as in WMAP7 (Hinshaw et al. 2013), assuming a spatially flat $\Lambda$CDM cosmology. Black holes are seeded at $M_{BH} = 10^5 M_{\odot}h^{-1}$ in all haloes once they reach $M_{\rm Halo} = 5 \times 10^{10} M_{\odot}h^{-1}$. As in Eagle, black hole growth is regulated via Bondi-Hoyle accretion, limited by the Eddington rate. Illustris operates two distinct feedback modes, although only one is effective at impacting star formation within galaxies. The first is a `quasar' mode which operates using the same general principle as the single mode in Eagle (outlined above). However, the energy injection is continuous, rather than bursty (see Sijacki et al. 2007, 2015). This mode is uncorrelated with quenching in the simulation. The second is a `radio' mode, which aims to simulate the effects of relativistic jets on the circum-galactic medium (CGM). At low accretion rates ($\chi_{\rm Edd} = \dot{M}_{BH} / \dot{M}_{\rm Edd} < 0.05$), once a black hole increases its mass by 15\% of its value, a bubble is seeded in the host galaxy's CGM, with energy $E_{\rm bubble} = \epsilon_m \epsilon_r \Delta M_{BH} c^2$, where $\epsilon_m$ represents the coupling efficiency of the mechanical feedback to the hot gas halo. Heating is induced in the CGM through $PdV$ work as the bubble expands. This mode is {\rm partially} effective at shutting down gas cooling from the CGM, and hence reducing star formation in the galaxy via starvation. However, as is now widely known, the jet bubbles also have the deleterious effect of completely vacating the CGM of gas in stark contrast to observations (e.g., Nelson et al. 2018; Pillepich et al. 2018).

\subsubsection{IllustrisTNG}

\noindent For IllustrisTNG, we utilize the {\small TNG-100-1} simulation (Nelson et al. 2018; Pillepich et al. 2018), which has an identical box size to our selected run in Illustris. IllustrisTNG adopts cosmological parameters from Planck Collaboration (2016). This simulation offers the best compromise between resolution and volume for our present study (see Piotrowska et al. 2022). TNG was run with an updated version of {\it \small AREPO}, extended to add magnetic fields to the implementation. Black holes are seeded at a higher mass of $M_{BH} = 8 \times 10^5 M_{\odot}h^{-1}$ in all haloes once they reach $M_{\rm Halo} = 5 \times 10^{10} M_{\odot}h^{-1}$. As in Eagle and Illustris, black hole growth is modeled sub-grid via Eddington limited Bondi-Hoyle accretion. The `quasar' mode feedback is left identical to Illustris, and it still has very little impact on star formation or quenching (see Weinberger et al. 2017). Alternatively, the `radio' mode feedback of Illustris (which was over-zealous in its removal of CGMs, though effective at quenching galaxies) is replaced with a new `kinetic' mode. When a black hole is accreting at a `low' level (defined relative to the black hole mass, see  Weinberger et al. 2017), energy injection is applied kinetically to a group of neighboring gas cells in a stochastic manner. This occurs at a threshold black hole mass of $M_{BH} \sim 10^8 M_\odot$, which is set partly by the sub-grid AGN feedback prescription and partly by relation to other evolving parameters in the simulation (see Zinger et al. 2020 for a full discussion on this). The change in kinetic energy of a gas cell is given by $\dot{E}_{\rm kinetic} = \epsilon_{k} \dot{M}_{BH}c^2$, where $\epsilon_{k}$ is the efficiency of energy transfer. The efficiency itself is set as a function of the gas density around the black hole (see Weinberger et al. 2017). Ultimately, a momentum kick is applied in a randomly chosen direction, such that (integrated over time) isotropy is preserved. Unlike the radio mode in Illustris, the kinetic mode in TNG impacts the ISM as well as the CGM. In slightly more detail, the TNG kinetic mode drives winds in the ISM as well as jet-like features in the CGM, which simultaneously adds turbulence to the ISM and increases the entropy (and hence cooling time) of the CGM. This occurs without removing significant quantities of gas from either, resolving the severe issues in Illustris (see Zinger et al. 2020: Piotrowska et al. 2022).

\subsubsection{Simulations Summary}

\noindent Eagle, Illustris, and IllustrisTNG represent three contemporary galaxy evolution models which all quench star formation in massive galaxies via AGN feedback. It is important to appreciate that, although very different in the details, all of the above AGN feedback models extract energy from around the black hole (ultimately some fraction of the accreted rest energy, $M_{BH} c^2$). Hence, the total feedback energy is directly proportional to black hole mass in each case, i.e. $E_{\rm feedback} \propto M_{BH}$. This suggests a way to test the entire paradigm of AGN feedback quenching in an essentially model independent manner (see Section 4.1, as well as Bluck et al. 2020a; Piotrowska et al. 2022). Nonetheless, there are many important differences between these three galaxy formation simulations in terms of quenching at a quantitative level (see, e.g., Donnari et al. 2021; Piotrowska et al. 2022).

\subsection{Observations}

\noindent We compare the predictions for AGN feedback driven quenching from simulations to the results from two observational galaxy surveys: the Sloan Digital Sky Survey - SDSS\footnote{SDSS Data Access: https://classic.sdss.org/dr7/access/} (Abazajian et al. 2009), and the Cosmic Near-Infrared Deep Extragalactic Survey - CANDELS\footnote{CANDELS Data Access: http://arcoiris.ucolick.org/candels/} (Grogin et al. 2011; Koekemoer et al. 2011). For CANDELS, we additionally utilize value added catalogs from Dimauro et al. (2018)\footnote{CANDELS VAC: https://mhuertascompany.weebly.com/data-releases-and-codes.html}. A thorough description of the parameter estimation of galaxies from these surveys is provided in Bluck et al. (2022). Here we give only a brief overview of the most important measurements used in this work.

\subsubsection{SDSS}

\noindent We take star formation rates (SFRs) for SDSS galaxies from Brinchmann et al. (2004), which are computed from emission lines where possible, or else from an empirical relationship between D4000 - sSFR for non-emission line galaxies (or systems with a significant AGN contribution). Fibre SFRs are converted to global SFRs utilizing the colors of galaxies outside of the aperture (see Brinchmann et al. 2004 for a full description). We take stellar masses from the Mendel et al. (2014) SDSS mass catalogs. We take galaxy size estimates from the Simard et al. (2011) morphological catalogs. Additionally, we utilize volume weights from Bluck et al. (2014) to simulate the statistical appearance of a volume complete sample, as appropriate for simulation comparison. Finally, we utilize the group catalogs of Yang et al. (2007, 2009) to separate central and satellite galaxies in the SDSS. Centrals are defined as the most massive galaxy in the dark matter halo, with isolated galaxies also counting as centrals for our purpose. Satellites are defined to be any other galaxy contained within the group (or cluster) halo. Extensive details on all of these measurements, as well as numerous checks on their reliability, are provided in Bluck et al. (2014, 2019, 2022). For our analysis, we select galaxies to be centrals, have $M_* > 10^9 M_{\odot}$, and reside within haloes with masses $M_{\rm halo} > 10^{11} M_{\odot}$ (which is identical to our simulations selection).

\newpage

\subsubsection{CANDELS}

\noindent We utilize photometric star formation rates; stellar masses of galaxies, bulges and disks; galaxy sizes; and rest-frame/ dust corrected UVJ colors from SED fitting of CANDELS galaxies provided in the value added catalogs of Dimauro et al. (2018). The SED fitting is performed utilizing the {\it \small FAST} code (Kriek et al. 2009), with stellar population synthesis models taken from Bruzual \& Charlot (2003), assuming a Chabrier IMF and Calzetti et al. (2000) extinction curve. Since CANDELS is a photometric only survey, unlike the SDSS, it is not possible to construct accurate central - satellite segregation, or group determination, due to the lack of precision in photometric redshift estimates compared to spectroscopic redshift measurements. As such, we analyze the full CANDELS galaxy sample, as opposed to focusing only on centrals. This is unlikely to be a significant problem since the vast majority of galaxies of these masses are predicted to be centrals at all epochs ($\sim$80\%, see, e.g., Henriques et al. 2015). Additionally, we select galaxies to have stellar masses $M_* > 10^9 M_{\odot}$ (in line with the simulations and the SDSS), but do not apply a halo mass cut (due to a lack of accurate halo estimation in this survey). We note that in the SDSS the impact of additionally applying a halo mass cut is very small ($< 5\%$ of the sample is affected and all results are invariant). Full details on the CANDELS measurements are given in the above references, and an extensive discussion on the reliability of this data is provided in Bluck et al. (2022).


\section{Methods}

\subsection{Identifying Quenched Systems Throughout Cosmic Time}

\begin{figure*}
\includegraphics[width=0.5\textwidth]{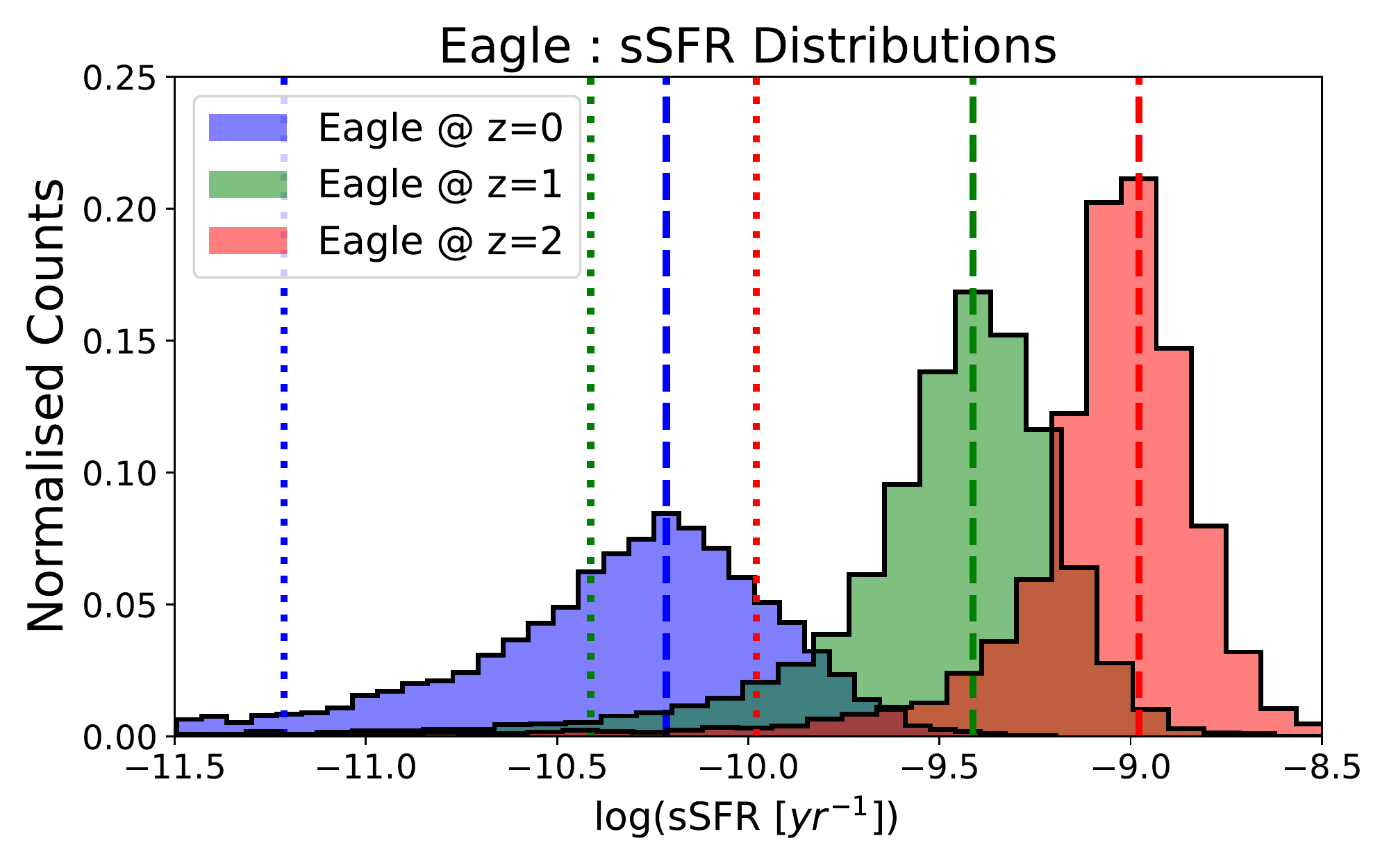}
\includegraphics[width=0.5\textwidth]{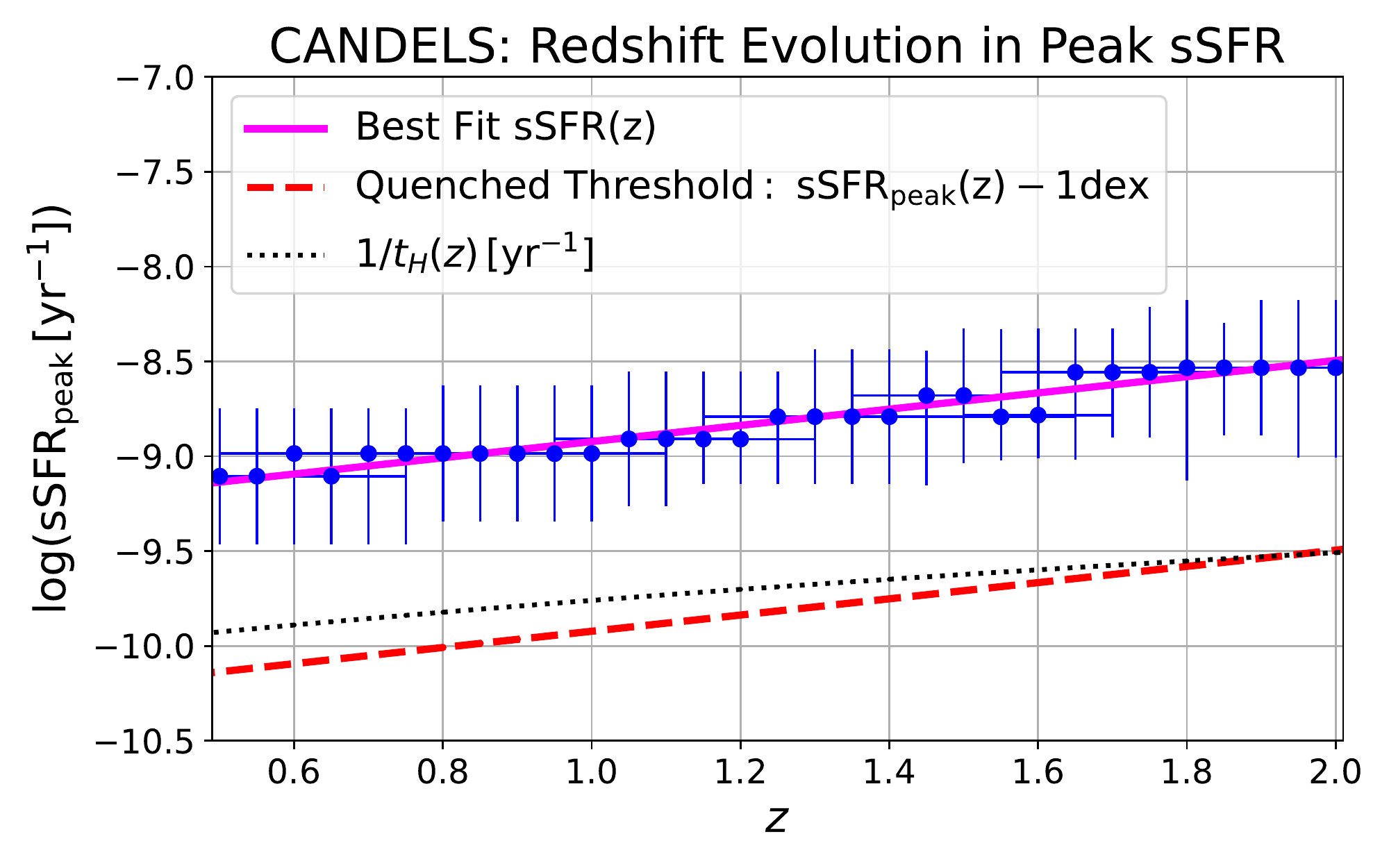}
\caption{Illustration of our method to select quenched and star forming galaxies in simulations and observations. {\it Left Panel: } sSFR distributions for three redshift snapshots from the Eagle simulation. The peak sSFR is indicated by dashed lines, and our adopted quenched threshold (at sSFR = sSFR$_{\rm peak}$ - 1dex) is indicated by dotted lines. {\it Right Panel: } Evolution in peak sSFR from z = 0.5 - 2 in CANDELS. Blue dots display the peak of the sSFR distribution, with $x$, $y$ error bars indicating the $z$-range applied and 1$\sigma$ dispersion in sSFR, respectively. A linear best fit to the peak sSFR evolution is shown in magenta, with our corresponding quenched threshold shown as a dashed red line. For comparison, the inverse of the Hubble time is displayed, which provides a reasonable quenching threshold at these epochs. }
\end{figure*}

\noindent In this paper we utilize a simple method to segregate actively star forming and quenched galaxies at multiple epochs. We apply the exact same method in simulations and observations to mitigate the potential for bias through the use of heterogeneous approaches. However, since the level of star formation in galaxies of a fixed stellar mass varies as a strong function of redshift (e.g. Madau \& Dickinson 2014), it is necessary to allow the method to reflect this.

Explicitly, we define quenched galaxies to be any system with a specific star formation rate:

\begin{equation}
\mathrm{sSFR}(z') < \mathrm{sSFR}_{\mathrm{peak}}(z=z') - 1 \, \mathrm{dex}
\end{equation}
  
\noindent where, sSFR$_{\mathrm{peak}}(z=z')$ indicates the peak (mode) of the sSFR distribution at a redshift equal to the galaxy in question (i.e. $z=z'$). Hence, quenched galaxies are defined to be forming stars at a rate less than one order of magnitude below the star forming peak at the redshift of each object (approximately 3$\sigma$ below the main sequence relation). In the simulations, all systems with SFR = 0 are defined as quenched and included in all analyses. Star forming systems are considered to be any galaxies with sSFR values higher than the quenched limit. This method is illustrated for one simulation (Eagle) in Fig. 1 (left-hand panel). Note that we do not apply a mass term to our sSFR limit unlike in some of our prior work (e.g., Bluck et al. 2014, 2016). The reason for this is two-fold. First, the impact of sSFR variation with stellar mass is weak except at the high mass end, where the majority of systems are quenched in any case. Hence, this does not impact our classification results in any significant manner (as we have checked). Second, we choose this cut to be in line with our prior work (Piotrowska et al. 2022), where this decision was made to maximize the objectivity of the quenching criteria between different models and observations, which have different mass dependencies. Finally, the curve in the sSFR - $M_*$ relation is likely due to high mass star forming systems having quenched cores. Thus, it remains an open question conceptually whether or not this ought to be accounted for (see, e.g.,  Bluck et al. 2020a).

However, the CANDELS data spans a wide range of redshifts. To combat this, we compute the peak sSFR of star forming systems at multiple epochs, and fit a simple redshift relation to this (as illustrated in Fig. 1, right-hand panel). Additionally for CANDELS, we also consider a check on the sSFR method utilizing UVJ colors (exactly as in Bluck et al. 2022). The results from these two approaches are essentially identical.

We have explored alternative definitions of quenching, including an offset from the redshift evolving star forming main sequence relation ($\Delta$SFR, which fully accounts for potential variations in mass on the quenching definition), different threshold cuts, and excluding green valley systems with intermediate levels of star formation. The results in this paper are completely stable to these alternative methods for identifying quenched systems.

\subsection{Random Forest Classification: Revealing Causality with Machine Learning}

\noindent In this paper, we utilize a machine learning approach for classifying star forming and quenched systems in simulated and observed galaxy surveys. More specifically, we adopt a Random Forest classifier from the powerful {\it ScikitLearn} Python package (Pedregosa et al. 2011)\footnote{ScikitLearn: https://scikit-learn.org/stable/}. This method is an ideal compromise between sophistication (the ability to account for highly non-linear and non-monotonic relationships between multiple parameters) and interpretability (the ability to extract meaningful insights from the trained classifier). Full details on this method are provided in Bluck et al. (2022),  Brownson et al. (2022), and Piotrowska et al. (2022), including numerous detailed tests. Of particular value for our present application is the capacity of the Random Forest classifier to separate causally related parameters from nuisance parameters in a classification problem (see Bluck et al. 2022 appendix B). Ultimately, this is achieved through controlling for all other parameters when ascertaining the importance of any given parameter to the classification problem at hand. Here we give a brief review of the most important aspects of our Random Forest classification method for this work. 

A Random Forest is a set of decision trees with differences between them ensured through bootstrapped random sampling. Two example decision trees from our analysis are shown in the Appendix (see Figs. 8 \& 9). In training, the criteria for each decision branch (in each decision tree) is set by selecting the optimal parameter (and threshold) to most effectively separate the classes of data, defined explicitly by the Gini impurity:

\begin{equation}
G(n) = 1 - \sum_i^{c=2} \big(p_i(n)^2\big)
\end{equation}

\noindent where $p_i(n)$ indicates the probability of randomly selecting class, $i$, at node, $n$. The summation is made over all classes (i.e. in this paper, over star forming and quenched systems). For example, in our application, the most useful observable (from a list containing black hole mass, stellar mass, halo mass etc.) is selected to maximize the accuracy with which quenched and star forming galaxies are separated at each branch. Then the optimal threshold in that variable is found to minimize the Gini impurity. The entire decision tree is filled out through iteration, until no further improvements are possible (or else a pre-defined threshold is reached). 

Once the Random Forest classifier is trained, one can extract how important each parameter is for solving the classification problem. The relative importance of a feature (i.e. parameter under investigation), $k$, is defined as:

\begin{equation}
I_R(k) = \frac{1}{N_{\mathrm{trees}}} \sum_{\mathrm{trees}} \bigg\{     \frac{\sum_{nk} N(n_k) \Delta G(n_k) } { \sum_{n} N(n) \Delta G(n)}     \bigg\}
\end{equation}

\noindent where the numerator in the above expression indicates the the improvement in Gini impurity ($\Delta G(n)$) weighted by the number of data points entering that node ($N(n)$), summed over all nodes which utilize feature parameter $k$ (hence the additional subscripts). The denominator provides a similar summation, but over all nodes, regardless of the features used. Hence, the ratio gives the relative importance of the variable $k$ for solving the classification problem (i.e. in this work, determining whether galaxies will be star forming or quenched) in a given decision tree. Finally, an average is taken over all tress in the Random Forest to give the final relative importance statistic of each training parameter. This is the key statistic we use in this paper, and from now on we will refer to it as the {\it quenching importance} (due to our specific science application).

To assess the uncertainties on the Random Forest classification, we run 25 independent training and testing runs of each analysis, taking the median importance as the final result and the standard deviation as the statistical error. We are careful to avoid over-fitting by testing the performance of the Random Forest classifier on unseen data. More specifically, we require a difference in performance (measured by the area under the true positive - false positive curve, AUC) of $\Delta$AUC $<$ 0.02 (see Teimoorinia et al. 2016; Bluck et al. 2022 for further details). The required (very) small difference in performance between the training and testing data sets prevents the classifier from learning pathological features in the training data which do not scale to unseen data.

Before training the random forest, all parameters are median subtracted and normalized by the interquartile range, to avoid differences in parameter variance or magnitude impacting the results. We train and test on a balanced sample of 50\% quiescent and 50\% star forming systems in each classification. This avoids weighting one population as more important than the other. We also reserve 50\% of the data for performance testing, which is unseen by the random forest classifier. This enables the rigorous over-fitting testing, described above. The final results are taken from the performance on the unseen data. The above methodology is essentially standard in the machine learning literature (see Teimoorinia et al. 2016; Bluck et al. 2019, 2020a,b and references therein for further discussion). It is important to apply these data preparation steps carefully to avoid biased results.\\


\section{Results \& Discussion}

\noindent In this section we apply our Random Forest classification technique (discussed in Section 3.2) to three cosmological hydrodynamical simulations, in order to identify the key observable signature of AGN feedback driven quenching across cosmic time (Section 4.1). Additionally, we apply the Random Forest technique to two observational wide-field galaxy surveys, in order to test the model predictions across a wide range of epochs (Section 4.2). We also discuss the implications of our results within the context of the literature.\\

\subsection{The Fundamental Signature of AGN Feedback Driven Quenching in Simulations across Cosmic Time}

\noindent In Fig. 2 we present the results from a series of Random Forest classification analyses applied to the problem of identifying quiescent galaxies in the IllustrisTNG, Eagle, and Illustris hydrodynamical simulations. Results are presented separately for each of the three cosmological simulations (shown in separate panels) and for five redshift snapshots spanning from $z$=0 to $z$=2 (as labelled by the legend in each panel). The parameters used to train the Random Forest are displayed along the x-axis, and the relative importance for quenching is displayed as the bar height (on the y-axis). Uncertainties on the quenching importance are estimated from the 1$\sigma$ dispersion of 25 parallel classification training and testing runs, applying random sub-sampling of the data.

For each simulation, and at every redshift snapshot considered, black hole mass is clearly identified as the most predictive parameter for classifying star forming and quenched galaxies. No other parameter reaches beyond $\sim$1/4 the relative importance of black hole mass at any epoch, and most parameters are of negligible importance for quenching at all redshifts. Hence, the key prediction of AGN feedback models is a clear dependence of galaxy quenching on black hole mass. It is important to highlight that this result is stable from cosmic noon to the present epoch, unanimously predicted by three independent simulations, and is (at least in principle) observationally testable.


\begin{figure*}
\begin{centering}
\includegraphics[width=0.75\textwidth]{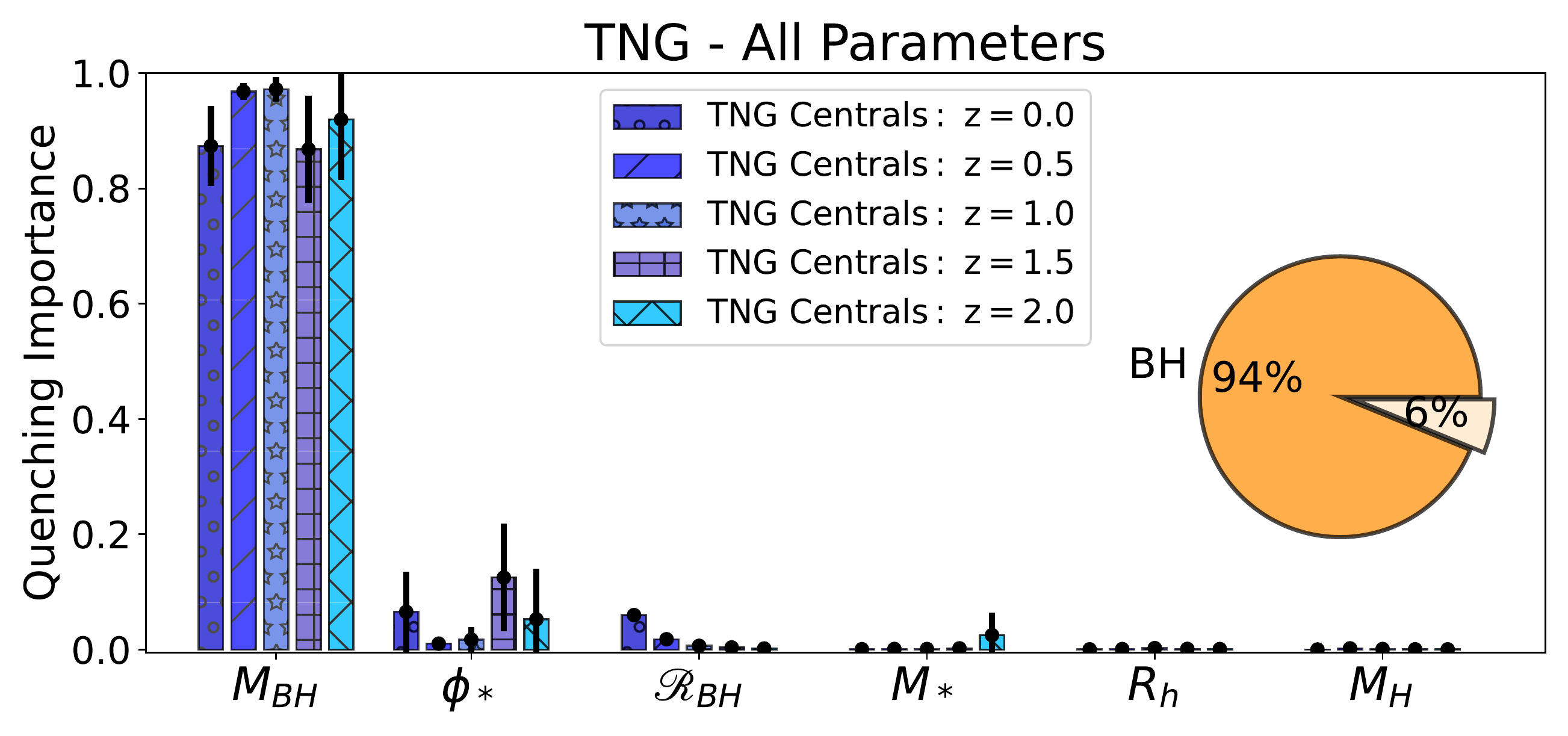}\\
\includegraphics[width=0.75\textwidth]{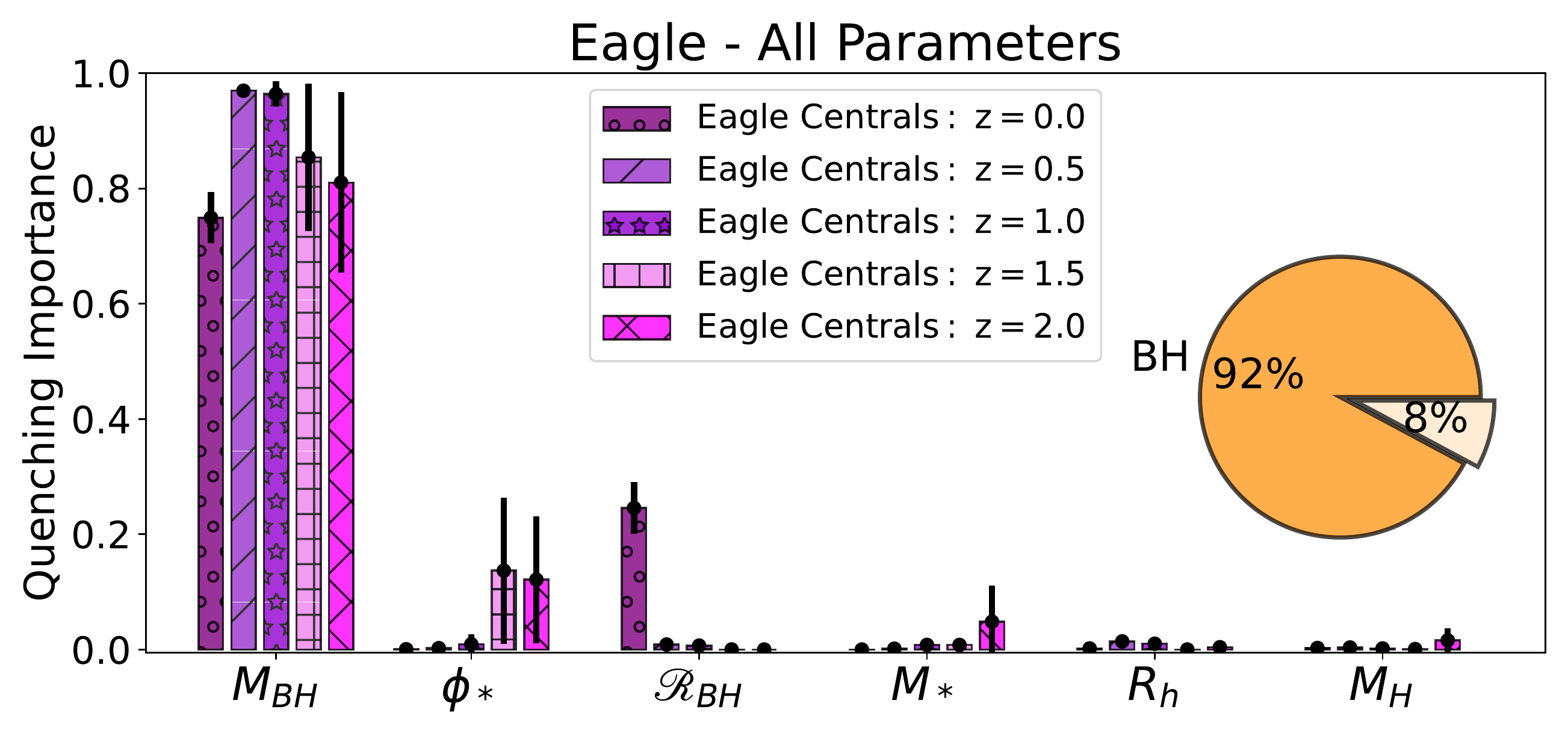}\\
\includegraphics[width=0.75\textwidth]{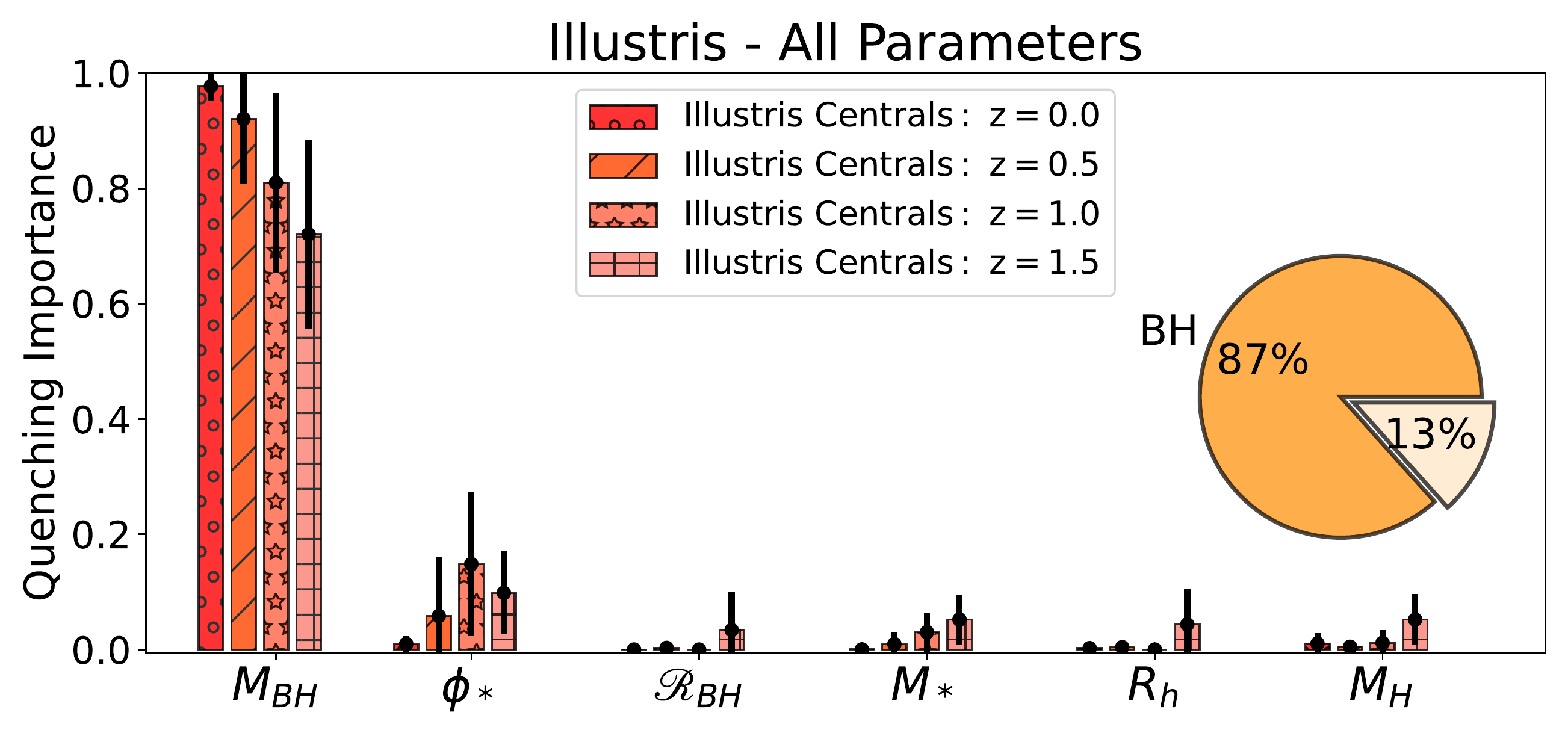}
\caption{Random Forest classification analysis to predict the existence of quiescent central galaxies in three cosmological simulations. The quenching importance of black hole mass ($M_{BH}$), the stellar potential ($\phi_*$), black hole accretion rate ($\mathscr{R}_{BH} \equiv dM_{BH}/dt$), stellar mass ($M_*$), stellar half mass radius ($R_h$), and dark matter halo mass ($M_H$) are displayed on the y-axis (as labelled from left-to-right along the x-axis of each panel). The results are shown separately for IllustrisTNG (top panel), Eagle (center panel), and Illustris (bottom panel), for a variety of snapshots spanning from cosmic noon ($z \sim 2$) to the present epoch (as shown on the legend of each panel). Note that the z=2 snapshot for Illustris is absent due to a lack of quenched galaxies at that epoch. Uncertainties on the quenching importances are given from bootstrapped random sampling. It is clear that black hole mass is overwhelmingly the most important parameter for predicting quiescence in central galaxies at all epochs (and for every simulation) considered here. A pie plot is displayed on each panel, which averages the quenching importances across all epochs, comparing two broad categories: BH-parameters (i.e. $M_{BH}$ and $\mathscr{R}_{BH}$); and non-BH-parameters (the remaining set). It is clear that non-BH parameters engender only a very minor improvement in predicting quiescence over BH-parameters alone.}
\end{centering}
\end{figure*}

It is especially important to note the utter lack of importance given to black hole accretion rate by the classifier in all simulations and at all redshifts studied\footnote{The only slight exception to this is for Eagle at the lowest redshift snapshot, where accretion rate performs as the second best parameter. Interestingly, this is the ‘exception that proves the rule'. The value of $\mathscr{R}_{BH}$ actually turns out to be that it {\it anti-correlates} with quenching. That is, higher accretion rates at low redshifts in Eagle are associated with more star forming systems: the opposite of catching AGN quenching in action!}. Given that the total bolometric AGN luminosity is directly proportional to accretion rate (i.e. $L_{\rm AGN} = \eta \, \mathscr{R}_{BH} \, c^2$, for an efficiency $\eta \sim 0.1$), this further implies a negligible importance to quenching of contemporaneous AGN luminosity (and hence also AGN identification). This result is in close agreement with Ward et al. (2022), who find that the most luminous AGN in simulations are found in the most gas rich (and hence star forming) systems. This demonstrates that there is expected to be no instantaneous link to cessation of star formation in contemporary models, despite AGN being used to quench galaxies in these simulations. Therefore, searching for evidence of AGN feedback by looking for a difference in (s)SFR between AGN and non-AGN systems is destined to failure, if AGN feedback operates as it is predicted to in modern hydrodynamical simulations. Nonetheless, a great number of studies have proceeded in just this manner (e.g., Nandra et al. 2007; Bundy et al. 2008; Georgakakis et al. 2008; Silverman et al. 2008; Hickox et al. 2009; Xue et al. 2010; Aird et al. 2012; Rosario et al. 2013; Heckman \& Best 2014; Trump et al. 2015; Ellison et al. 2016; Shimizo et al. 2017; Florez et al. 2020; and many more). Consequently, one must be cautious in the interpretation of these prior results. Indeed, from our analysis of hydrodynamical simulations, a lack of clear connection between AGN luminosity and quiescence is actually a feature of the modern AGN feedback quenching paradigm, not evidence against it. 

We speculate that the above result reflects the importance of preventative feedback in these models. In such a scenario, it is the energy released over long periods from AGN which really matters for quenching (which is clearly traced by $M_{BH}$ not  $\mathscr{R}_{BH}$). Consequently, quiescence may emerge as a long term consequence of AGN heating preventing gas cooling and accretion from the CGM into massive galaxies, ultimately starving the system of fuel needed for further star formation. This mode of operation simultaneously resolves the cooling problem in high mass groups and clusters (e.g., Fabian 2012), explains the low cosmological star formation efficiency (e.g., Fukugita \& Peebles 2004), as well as accounting for the observed demographics of star forming and quenched galaxies (e.g., Peng et al. 2010; Bluck et al. 2014). Instantaneous feedback may still trigger quenching (e.g., Zinger et al. 2020; Terrazas et al. 2020), but without long term heating of the CGM, star formation will inevitably reignite within massive galaxies, due to gas cooling and condensation into the system, removing the galaxy from our quenched sample. Since our machine learning classification analysis is sensitive to what is fundamentally different between quenched and star forming systems, our results pick up on the long term preventative mode rather than the instantaneous trigger (by design).

Additionally in Fig. 2, we present pie plot insets showing the results from a simplified analysis, averaging over all epochs in the simulations, and separating parameters coarsely by whether they are black hole related (i.e., $M_{BH}$ \& $\mathscr{R}_{BH}$) or non-black hole related (i.e., $M_*$, $R_h$, $M_H$, $\phi_*$). It is clear that BH-parameters dominate the predictive power in the Random Forest classification. Within this set, as discussed above, it is black hole mass not accretion rate which really matters for quenching. The rest of the parameters are all of very little importance for predicting quiescence, offering on average $<$15\% of the total quenching importance. For straightforward theoretical reasons (see Bluck et al. 2020a; Piotrowska et al. 2022), stellar mass is closely related to the total energy released from supernovae, and halo mass is closely related to the total energy released from virial shocks. Hence, the simulations predict that these important processes within galaxies have a negligible impact on the quenching of centrals across cosmic time. Again, this is a clear prediction of contemporary AGN feedback driven quenching models, which can be observationally tested (in principle at least). At low redshifts, the models clearly pass this test (see Piotrowska et al. 2022), yet it remains to be seen whether this is so at higher redshifts or not.

It is important to appreciate that even though Eagle, Illustris, and IllustrisTNG utilize very different sub-grid prescriptions for AGN feedback (including different processes at a fundamental level, e.g. CGM jet bubbles vs. SMBH-driven winds into the ISM; kinetic vs. thermal energy injection), the fundamental qualitative quenching predictions are identical between these simulations at all epochs. The reason for this is that black hole mass traces the total energy released from black hole accretion in a model independent manner (see Soltan et al. 1982; Silk \& Rees 1998; Bluck et al. 2020a; Terrazas et al. 2020; Zinger et al. 2020; Piotrowska et al. 2022). As such, these predictions are fundamental to any quenching process with energy originating from the central black hole. Note that this is true even though there are many quantitative differences between these simulations in terms of quenching (see, e.g., Donnari et al. 2021; Piotrowska et al. 2022, as well as Appendix A1). This is a profound point for two reasons. First, variation in the resolution, mechanism of AGN - CGM energy/ momentum transfer, sub-grid recipe, hydo-solver, or volume size of simulations are unable to change the dependence of quiescence on black hole mass without fundamentally abandoning AGN as the cause of quenching. Second, due to the first point, this enables an observational test of the entire paradigm of AGN feedback driven quenching, not just one specific instantiation of it. It is to this we turn in the next sub-section.\\

\subsection{Observational Tests of AGN Feedback Driven Quenching}

\noindent Conceptually, it is straightforward to test the AGN feedback paradigm - simply measure the parameters included in Fig. 2 for a representative sample of galaxies at each epoch and run the Random Forest classification. The problem is that many of these parameters, though observable in principle, are difficult to measure in practice. This is especially the case for black hole mass, i.e. the key observable in AGN feedback driven quenching (see Fig. 1). At low redshifts, we were able to leverage the power of statistical calibrations, e.g. the $M_{BH} - \sigma$ relation, in order to make progress (see Piotrowska et al. 2022). Yet, at redshifts beyond $z \sim 0.2$, even the proxy variables needed for accurate statistical calibrations of black hole mass (e.g. central velocity dispersion) are infrequently measured. This critical issue will be resolved in the coming years through dedicated spectroscopic surveys targeting intermediate-to-high redshifts, most notably with VLT-MOONS and JWST. Consequently, a robust test to the multi-epoch quenching predictions from hydrodynamical simulations (discussed in the previous sub-section) can soon be made.

Although not ideal for the purpose of constraining black hole mass, large photometric surveys exist with HST rest-frame optical imaging at the epochs from $z = 0.5 - 2$, the largest of which is HST-CANDELS (Grogin et al. 2011; Koekemoer et al. 2022). As such, it is interesting to attempt to find a useful proxy for black hole mass in the simulations, which is relatively easy to constrain in photometric data. 

In lieu of a direct kinematic measurement of black hole mass, or a measurement of central velocity dispersion, one can still make progress by making several assumptions and leveraging the power of the virial theorem. The gravitational potential ($\phi_g$) for a system with dynamical mass, $M_{\rm dyn}$, and gravitational radius, $R_{\rm grav.}$, is given by:

\begin{equation}
\phi_g \equiv  \frac{-GM_{\rm dyn}}{R_{\rm grav.}}    \sim   \frac{M_{\rm dyn}}{ R_{\rm grav.}}   \sim   \frac{M_*}{R_{h}}  \equiv  \phi_*
\end{equation}

\noindent where we make a series of assumptions intended to progressively simplify the measurements needed from left-to-right in eq. 4 above, whilst preserving a strong relation to the gravitational potential. Note that the use of the tilda is intended to specify ‘scales with' rather than ‘of the order of', since the latter is obviously not true when changing units. In the first step, we drop the constants (which are irrelevant for classification performance). In the second step, we approximate the dynamical mass with the stellar mass, and the gravitational radius with the half mass radius. Within $\sim 2-3\,R_h$, this is a good approximation for most high mass systems, although it is clearly imperfect (it ignores the contribution of gas and dark matter, as well as the structure of the mass distribution).

For a virialised system, the gravitational potential is related to the mean square velocity ($v_{\rm rms}$) as follows:

\begin{equation}
\phi_g = v_{\rm rms}^2 \sim \sigma^2 \sim M_{BH}^{1/2}
\end{equation}

\noindent where, for a pressure supported system, $v_{\rm rms} \sim \sigma$ (hence the second step). Even for rotationally supported systems there is a strong relationship between $\sigma$ and $v_{\rm rms}$ (e.g., Brownson et al. 2022), although the tightness of the relation reduces. To arrive at the final step in eq. 5 above, we use the fact that black hole mass is empirically determined to scale with velocity dispersion to approximately the fourth power (e.g., Saglia et al. 2016). Hence, we have found an approximate way to estimate supermassive black hole mass in photometric data. Explicitly,

\begin{equation}
M_{BH} \sim \phi_*^2
\end{equation}

\noindent Or, in other words, there is at least some logic in suspecting that $\phi_*$ (which we can measure in pure photometric data) may trace $M_{\rm BH}$. Yet, the above argument is little more than a reasoned guess; the proof lies in the simulations. It is important to appreciate that $\phi_*$ is not the only correlator to black hole mass that one can expect from simple arguments of this type. For instance, bulge mass (e.g., Bluck et al. 2014, 2022), central density (e.g., Cheung et al. 2012; Fang et al. 2013) and light concentration (e.g., Wuyts et al. 2011) are all also expected to correlate strongly with central velocity dispersion and hence black hole mass. Our purpose in utilizing $\phi_*$ is purely for convenience - it is easy to measure in both the observational and simulated data sets used in this work, without requiring significant further data processing.


\begin{figure*}
\begin{centering}
\includegraphics[width=0.75\textwidth]{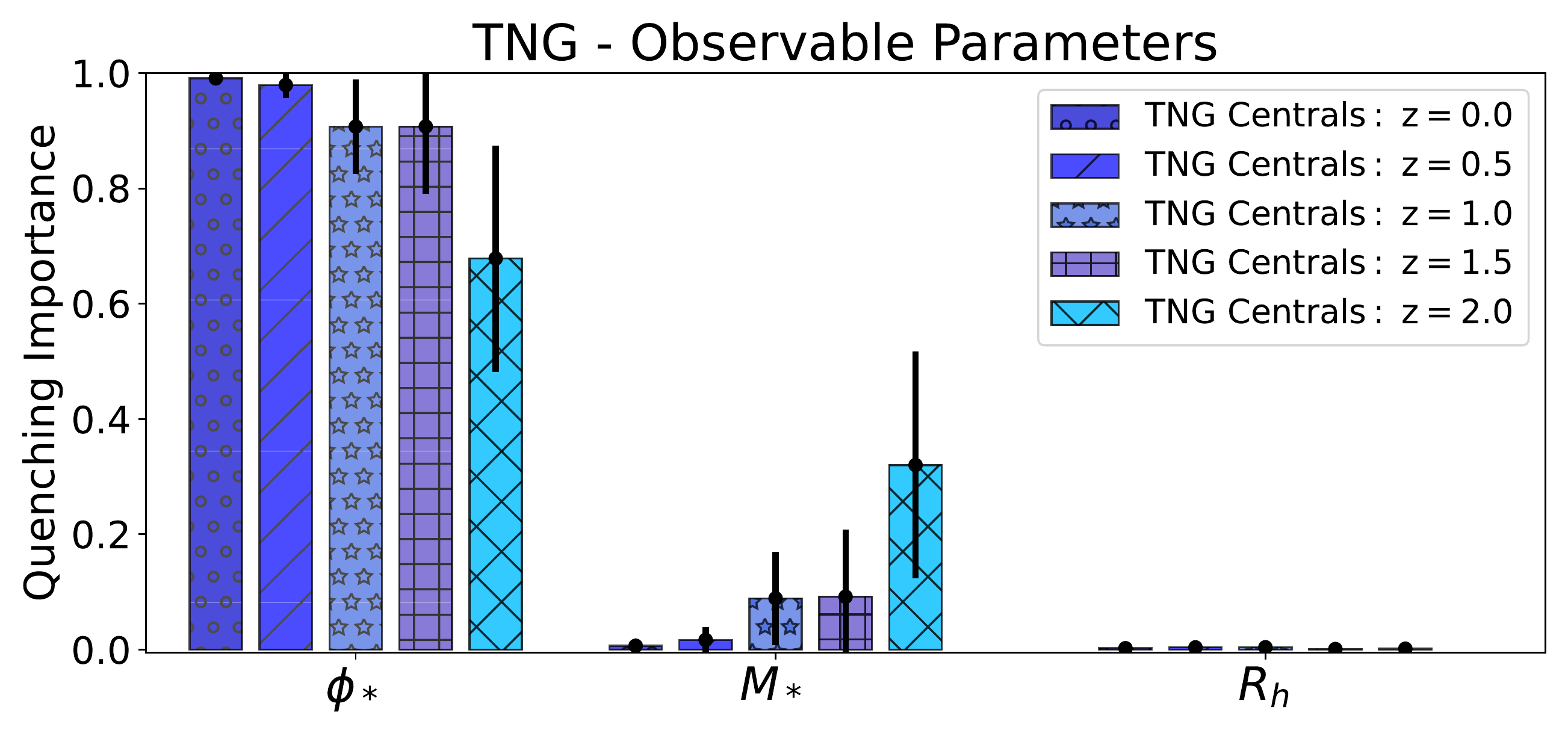}\\
\includegraphics[width=0.75\textwidth]{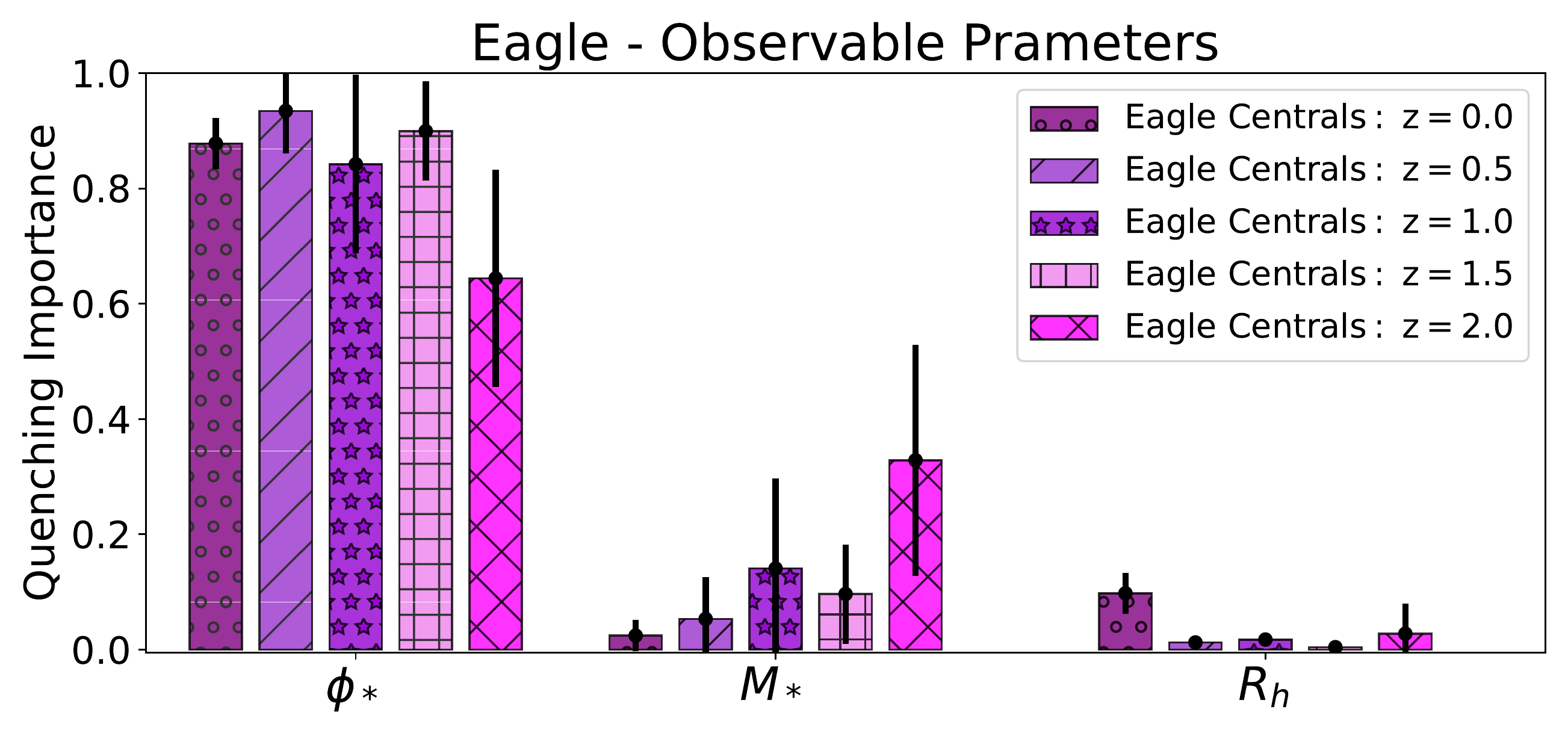}\\
\includegraphics[width=0.75\textwidth]{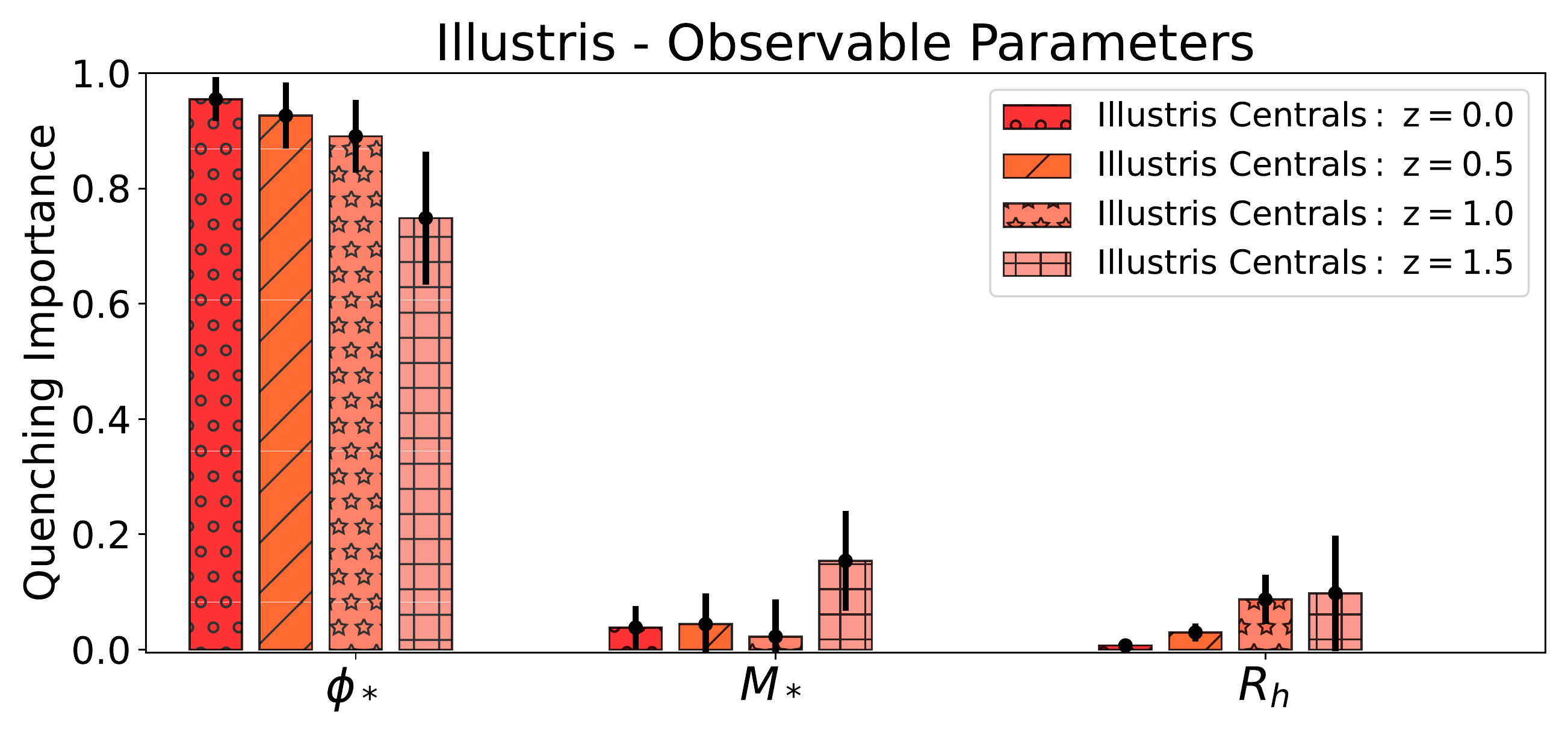}
\caption{Identical in structure to Fig. 2, except that fewer parameters are used to train the Random Forest classifier. Here we focus on stellar mass, half mass radius, and the stellar potential. These parameters are relatively straightforward to measure in extant wide-field galaxy surveys. Consequently, they offer an opportunity to test the predictions from the simulations. The stellar potential is clearly identified as the most important predictor of quenching, in lieu of black hole mass.}
\end{centering}
\end{figure*}


\begin{figure*}
\begin{centering}
\includegraphics[width=0.9\textwidth]{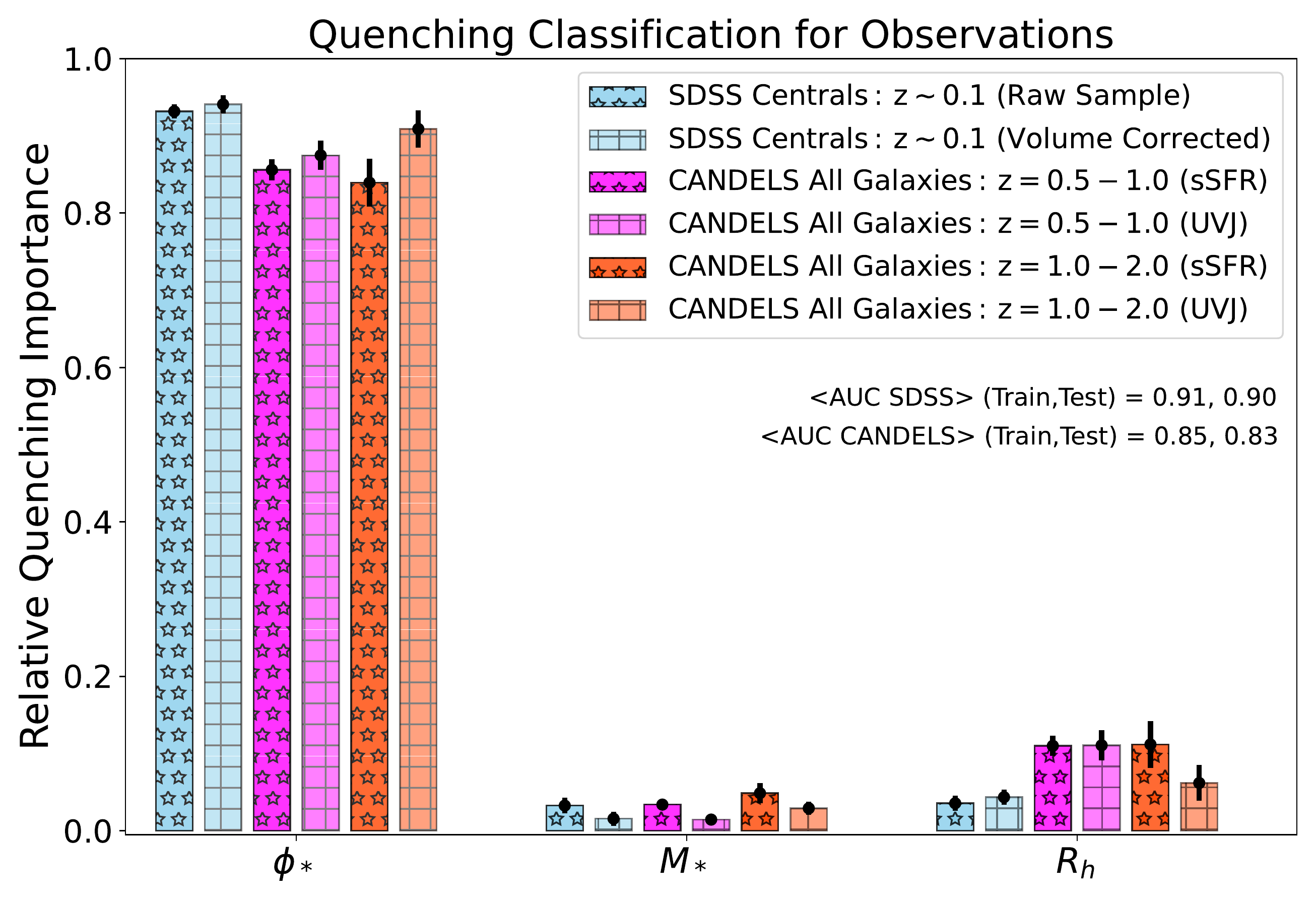}
\caption{Random Forest classification analysis to predict the existence of quiescent galaxies in observations. The relative quenching importance of the stellar potential ($\phi_*$), stellar mass ($M_*$), and rest-frame visible half light radius ($R_h$) are shown on the y-axis (as labelled by the x-axis). Uncertainties are given by boot strapped random sampling. At low redshifts (blue colored bars) we display results for the SDSS, using both raw galaxy counts and a $1/V_{\rm max}$ volume corrected sample. At higher redshifts we utilize data from the CANDELS survey, shown separately for intermediate redshifts (displayed in pink) and high redshifts (displayed in red). For the CANDELS data we consider both the sSFR approach (utilized throughout this paper) and a UVJ color based approach to identify quiescence. At all epochs (and for all methodological sub-samples), the stellar potential is clearly identified as the most predictive parameter for identifying quiescent systems. This is precisely as predicted by all three cosmological simulations studied in this paper (compare to Fig. 3). }
\end{centering}
\end{figure*}

In Fig. 3 we present a series of Random Forest classification analyses to predict the existence of quenched galaxies within the three simulations studied in this paper, exactly as in Fig. 2. However, here we restrict the parameters used to train the classifier to: stellar mass ($M_*$), half mass radius ($R_{h}$), and the stellar potential ($\phi_* = M_* / R_{h}$). All of these parameters can be estimated in a straightforward manner in extant photometric observations. It is clear from Fig. 3 that in lieu of black hole mass, $\phi_*$ is predicted to be the most important parameter for constraining quenching, at all epochs and for all simulations. This is as expected from the simple argument sketched above. More physically, in the simulations, this result emerges as a consequence of supermassive black holes forming and growing in the dense gravitational potentials at the center of galaxies (e.g. Sijacki et al. 2007, 2015; Weinberger 2018).

In Fig. 4 we test the above prediction from hydrodynamical simulations in observational data. At low redshifts we analyze the SDSS (blue bars), and at higher redshifts we analyze CANDELS (shown in pink for intermediate redshifts, and in red for high redshifts). Additionally, for the SDSS, we consider two subsamples: raw data counts and a volume corrected sample. The latter mimics a volume complete sample of galaxies, testing whether Malmquist bias could impact the results. For CANDELS, we consider the standard sSFR method for identifying quenched systems (used throughout this paper), as well as a UVJ color method (as in Bluck et al. 2022). Since in CANDELS we do not have access to spectroscopic star formation rates, we utilize this extra test to ensure that the results of the Random Forest classification are stable to alternative methods for defining quenching. 

For all redshifts, galaxy surveys, and methodological choices, in Fig. 4 we find that the stellar potential is by far the most predictive parameter of quenching in observations, compared to stellar mass and galaxy size\footnote{Note that for the observational data we utilize half light radius instead of half mass radius, choosing the reddest waveband available (with high quality data) in order to minimize the systematic offset. Explicitly, we utilize $r$-band in the SDSS and $H$-band in CANDELS, which both trace the rest-frame optical at the redshifts probed by each survey. This could be a source of {\it difference} between observations and simulations, but in no way could this ensure a greater similarity.}. Hence, there is remarkable agreement between the simulation predictions and the observational results. This is a clear success of the models, and by extension the paradigm of AGN feedback driven quenching.

The importance of the stellar potential for predicting quiescence is also consistent with numerous other observational works investigating bulge mass, central mass density, central velocity dispersion, and light concentration as phenomenological drivers of galaxy quenching (see Bell et al. 2008, 2012; Wuyts et al. 2011; Wake et al. 2012; Cheung et al. 2012; Fang et al. 2013; Omand et al. 2014; Bluck et al. 2014, 2016, 2020a,b, 2022; Piotrowska et al. 2022; Brownson et al. 2022; Varma et al. 2022). The comparison with modern cosmological simulations performed in this paper establishes a plausible physical origin to these prior results. That is, AGN feedback driven quenching is best predicted by the fossil record of past accretion history (i.e. black hole mass), which is itself highly correlated with the local potential well in which the black hole forms (approximately given by $\phi_*$). Naturally, the local potential also scales with central mass density, stellar light concentration, mass of the central bulge component, and central velocity dispersion, hence explaining these prior results as a plausible consequence of historic AGN feedback. 

Of course, whilst the above results are consistent with the fundamental prediction from AGN feedback models, it remains possible that other mechanisms could yield a tight dependence of quenching on the stellar potential as well. To fully rule out this possibility, dynamical measurements of the masses of supermassive black holes are needed on a large scale at intermediate-to-high redshifts.

\newpage


\section{Summary}

\noindent In this paper we apply Random Forest classification to three cosmological hydrodynamical simulations (Eagle, Illustris, and IllustrisTNG) and two wide-field galaxy surveys (SDSS \& CANDELS), identifying the optimal parameters for predicting whether central galaxies will be star forming or quenched. This work expands on our initial low redshift analysis in Piotrowska et al. (2022) by probing the dependence of quenching on physical parameters out to cosmic noon. Throughout this paper we focus on galaxies in the stellar mass range $9 < \log(M_*/M_{\odot}) < 12$.
\\\\
Our principal results are as follows:

\begin{enumerate}

\item Supermassive black hole mass is identified as the most important parameter for predicting quiescence in all three simulations, and at every redshift snapshot analyzed (spanning $\sim$10\,Gyr of cosmic history). This is the key testable prediction from models of AGN feedback driven quenching. 

\item Perhaps surprisingly, supermassive black hole accretion rate is not constraining of galaxy quenching at any epoch, according to the simulations. This further implies that AGN luminosity is not predicted to correlate strongly with galaxy quiescence. Instead, it is the {\it integrated} effect of historic AGN feedback which leaves a clear signature on the galaxy population (traced by $M_{BH}$ not $L_{\rm AGN}$).

\item In lieu of a measurement of black hole mass, the stellar potential ($\phi_* \sim M_* / R_h$) is predicted to act as a good proxy in simulations, outperforming all other variables for predicting quenching.

\item We confirm the above prediction from simulations in observations, utilizing the SDSS at low redshifts and CANDELS at intermediate-to-high redshifts.

\end{enumerate}

\noindent Hence, we find a remarkable consistency between simulations which utilize AGN feedback to quench galaxies, and multi-epoch observations. Further tests on the Random Forest results, which confirm the above conclusions, are provided in the Appendix. Ultimately, the key to testing the paradigm of AGN quenching is to study the fossil record of past accretion (i.e., black hole mass) or its best available proxy, not a proxy of accretion rate (such as AGN luminosity). More precise observational tests of the key AGN quenching prediction from simulations will become feasible with VLT-MOONS and JWST observations in the coming years.\\\\

\newpage

\section*{Acknowledgements}

\noindent AB acknowledges a faculty start-up grant at the Florida International University. JMP acknowledges funding from the MERAC Foundation. RM acknowledges ERC Advanced Grant 695671: ‘QUENCH’, support from the Science and Technology Facilities Council (STFC), and support from a Royal Society Research Professorship. 

We thank the anonymous referee for a highly positive and constructive report, which has helped to significantly improve this publication. We are grateful to the Illustris, Eagle, and IllustrisTNG teams for making their simulations public. We are especially grateful to Paul Torrey, Joop Schaye, Dylan Nelson, Annalisa Pillepich, Lars Hernquist, and Rob Crain for many stimulating and enlightening conversations on the simulations used in this work. We thank the SDSS and CANDELS teams for making their observational surveys public. We are especially grateful to Luc Simard, Sara Ellison, Trevor Mendel, Marc Huertas-Company, and Paula Dimauro for much advise and support with these data products. We also thank Yingjie Peng, Emma Curtis-Lake, Gareth Jones, Stephen Eales, Mirko Curti, Sara Ellison and Christopher J. Conselice for many fruitful and engaging conversations on this work.\\

\section*{Data Availability}

\noindent All of the data used in this work is publicly available, see the following links for access:

\begin{enumerate}

\item EAGLE: virgodb.dur.ac.uk/ 

\item Illustris: www.illustris-project.org/

\item IllustrisTNG: https://www.tng-project.org/

\item SDSS DR7: https://classic.sdss.org/dr7/access/

\item SDSS MPA-JHU release of spectrum measurements: www.mpa.mpa-garching.mpg.de/SDSS/DR7/

\item SDSS NYU Galaxy Value Added Catalogue: sdss.physics.nyu.edu/vagc/

\item SDSS Morphological Catalogues I: doi.org/10.1088/0067-0049/196/1/11 

\item SDSS Morphological Catalog II: doi.org//10.26093/cds/vizier.22100003

\item SDSS Group Catalogue: gax.sjtu.edu.cn/data/Group.html

\item CANDELS Main Release: http://arcoiris.ucolick.org/candels/

\item CANDELS VAC: https://mhuertascompany.weebly.com/data-releases-and-codes.html

\end{enumerate}

\vspace{1cm}


\bibliography{sample63}{}
\bibliographystyle{aasjournal}

\appendix

\section{Tests on the Random Forest Results}

\noindent In this appendix we present a number of detailed tests on the Random Forest classification results from the main body of this paper. All of the conclusions from the Random Forest analyses are consistent with the conclusions drawn from the alternative methods outlined here. As such, this appendix supports the main results of this paper.

\subsection{Simulations: Area Statistics at z = 1}

\noindent The results from Random Forest classification in the simulations are extremely clear: black hole mass is predicted to be the most important parameter regulating central galaxy quenching. In this part of the appendix, we show that this conclusion can be arrived at through a different analysis, which has some advantages over the Random Forest approach, and some disadvantages. Consequently, this provides a useful test on one of the main conclusions from this paper.

We adopt our area statistics method (outlined in detail in Bluck et al. 2016, 2020a) to assess which parameters engender the most significant impact on the quenched fraction, at fixed values of the other parameters. As in the main body of this paper, we select central galaxies consistently at $9 < \log(M_*/M_{\odot}) < 12$ and $M_{\rm Halo} > 10^{11} M_{\odot}$. Most of the variables analyzed in the area statistics plots have similar (though not identical) ranges between the simulations. The one exception is black hole accretion rate, where the lower limit varies significantly between the simulations. To combat this, we present all panels with the same range (to aid in comparison) but compute the area statistics only from the minimum to maximum value of each parameter in the source catalogs at this epoch, to avoid biasing this statistic (as in Bluck et al. 2016).

One major advantage of the area statistics approach is that it is highly visual and intuitive. On the other hand, this method only allows one variable to be controlled for at a time. Moreover, the area statistics approach is far less efficient than the Random Forest classification. Due to the latter issue, we restrict our analysis here to just one redshift snapshot (z = 1) and limit the parameters under consideration to: black hole mass, black hole accretion rate, stellar mass, and halo mass. 

In Fig. 5 we show the results from the area statistics approach for TNG galaxies at z = 1. The top set of panels shows the results for a balanced sample (50\% quenched; 50\% star forming), as utilized in the Random Forest classification. Additionally, we also present the results for the full volume complete sample, shown as the bottom set of panels. It is clear that the $f_Q - M_{BH}$ relation is by far the tightest of all of the relations considered in Fig. 5. This is true for both a balanced and complete parent sample. Consequently, black hole mass engenders a much larger impact on the fraction of quenched galaxies in the TNG simulation (at fixed other parameters) than any other parameter (at fixed black hole mass). Therefore, black hole mass is clearly established as the most fundamental parameter governing central galaxy quenching. This is in precise accord with the results from the top panel of Fig. 2.

In Figs. 6 \& 7 we repeat the analysis for Eagle and Illustris (respectively) also selected at z = 1. Again, it is clear that the $f_Q - M_{BH}$ relation is the tightest and steepest of all of the relations. This confirms the results from the center and bottom panels of Fig. 3. Additionally, we have explored the area statistics approach for all other redshift bins, and the full list of parameters. Black hole mass is clearly established as the most fundamental parameter in all of these extra tests as well.

As a result of the above analysis, it is evidently possible to arrive at the main conclusion of this paper for simulations through area statistics. However, this is highly inefficient, as it would require $\sim$20 pages of plots, rather than a single figure. More importantly, the Random Forest solves the full classification problem, controlling for all nuisance variables, and accounting for both the steepness and tightness of the quenched fraction relationships simultaneously (see Bluck et al. 2022). Hence, Random Forest classification is a much superior method to area statistics both conceptually and practically. Nevertheless, the latter does offer further insight into why the Random Forest yields the results it gives, as well as providing a simple visual check on the main conclusions.

Another advantage of the area statistics approach is that it allows us to probe some of the details of the process of quenching in the simulations. Although not the primary focus of this work, it is clear from Figs. 5 - 7 that central galaxies quench much more rapidly as a function of black hole mass (almost as a step function) in IllustrisTNG, but quench much more gradually as a function of black hole mass in both Eagle and Illustris. Additionally, the quenching threshold in black hole mass is over an order of magnitude higher in Illustris and IllustrisTNG, compared to Eagle. In this work we have not attempted to estimate black hole masses for observed galaxies, so we are not in a position to judge which is right. This notwithstanding, in Piotrowska et al. (2022) we do make a careful comparison of quenching fractions between these three simulations and low-z observations. No single simulation is an ideal march for the observational data, but IllustrisTNG is the closest. This indicates that, at least at present, a step function quenching threshold in black hole mass cannot be ruled out by the observational data.

\subsection{Example Decision Trees}

\noindent In Fig. 8 we present an example of one randomly selected decision tree from the Random Forest quenching classification of the Eagle simulation at z = 1. This illustrates the structure of decision trees. Black hole mass is selected as the parameter for the first split in the data. The reason for this is that a cut in black hole mass engenders the largest reduction in Gini impurity, over any other parameter. Consequently, black hole mass acquires the highest weight, and largest difference in Gini coefficient, leading to the highest importance in solving the classification problem (see eq. 3). Subsequent decision thresholds impact a smaller fraction of the data and have in general a lower reduction in Gini Impurity. Thus, it is crucial which parameter is chosen first by the classifier. The stability of this choice is assured by exploring 250 random generations of the parent sample, which is the fundamental advantage of a Random Forest over a single decision tree (see, e.g., Pedregosa et al. 2011; Bluck et al. 2022a).

In Fig. 9 we show an alternative decision tree from a classification analysis of Eagle at z = 1, utilizing a volume complete (as opposed to balanced) sample. Once again, black hole mass is chosen as the first parameter in the decision tree, engendering a high importance of this parameter. However, in this example, the sample is overwhelmingly star forming (due to the steepness of the mass function) and hence the classifier has to probe deeper to effectively separate the data. This is not the preferred approach for using machine learning in classification for various technical reasons (see, e.g., Teimoorinia et al. 2016). Nonetheless, this test clearly establishes the stability of the main results to the manner in which the data is presented to the classifier.

\subsection{Predicted $M_{BH} - \phi_*$ Dependence}

\noindent In Fig. 10 we show the predicted $M_{BH} - \phi_*$ relations from TNG (left panels), Eagle (center panels), and Illustris (right panels). It is clear that all three simulations predict that there should be a tight dependence of black hole mass on the stellar potential. This provides further insight into why $\phi_*$ is chosen by the Random Forest classifier in lieu of black hole mass (see Fig. 3). We can leverage this theoretical result to test the quenching predictions of these simulations in extant photometric observations.

In the SDSS, where we have the most complete suite of measurements on observational galaxy parameters, we have additionally made a number of extra tests. We find that  $\phi_*$ is second only to central velocity dispersion for predicting quiescence, comfortably beating bulge mass, total stellar mass, halo mass, B/T morphology, environmental parameters, and disk mass. However, the real value of  $\phi_*$ for our analysis is that it can be straightforwardly estimated in extant catalogs without further data processing, in both simulations and observations up to z = 2. Given that the simulations unanimously predict that this parameter will act as a reasonable proxy for $M_{BH}$, this is sufficient for our present analysis.

\subsection{Observations: Area Statistics in the SDSS \& CANDELS}

\noindent In Fig. 11 we apply our area statistics approach to the SDSS sample of central galaxies at low redshifts. It is clear that the $f_Q - \phi_*$ relation is by far the tightest of the set. This is exactly as predicted by all three simulations (see Fig. 3 \& 10). In Fig. 12 we apply the area statistics technique to the full CANDELS galaxy sample at z = 0.5 - 2. Again, the $f_Q - \phi_*$ relation is found to be the tightest (and steepest) of the observed relations, exactly as predicted by the simulations. This confirms the results from the Random Forest classification (shown in Fig. 4 in the main body of the paper). We have additionally tested the use of volume weighting, restricting to a balanced sample, including more redshifts cuts, and utilizing alternative quenching definitions (and thresholds) on the observational results. The results from this paper are extremely stable to these alternative analysis choices.



\begin{figure*}
\includegraphics[width=0.95\textwidth]{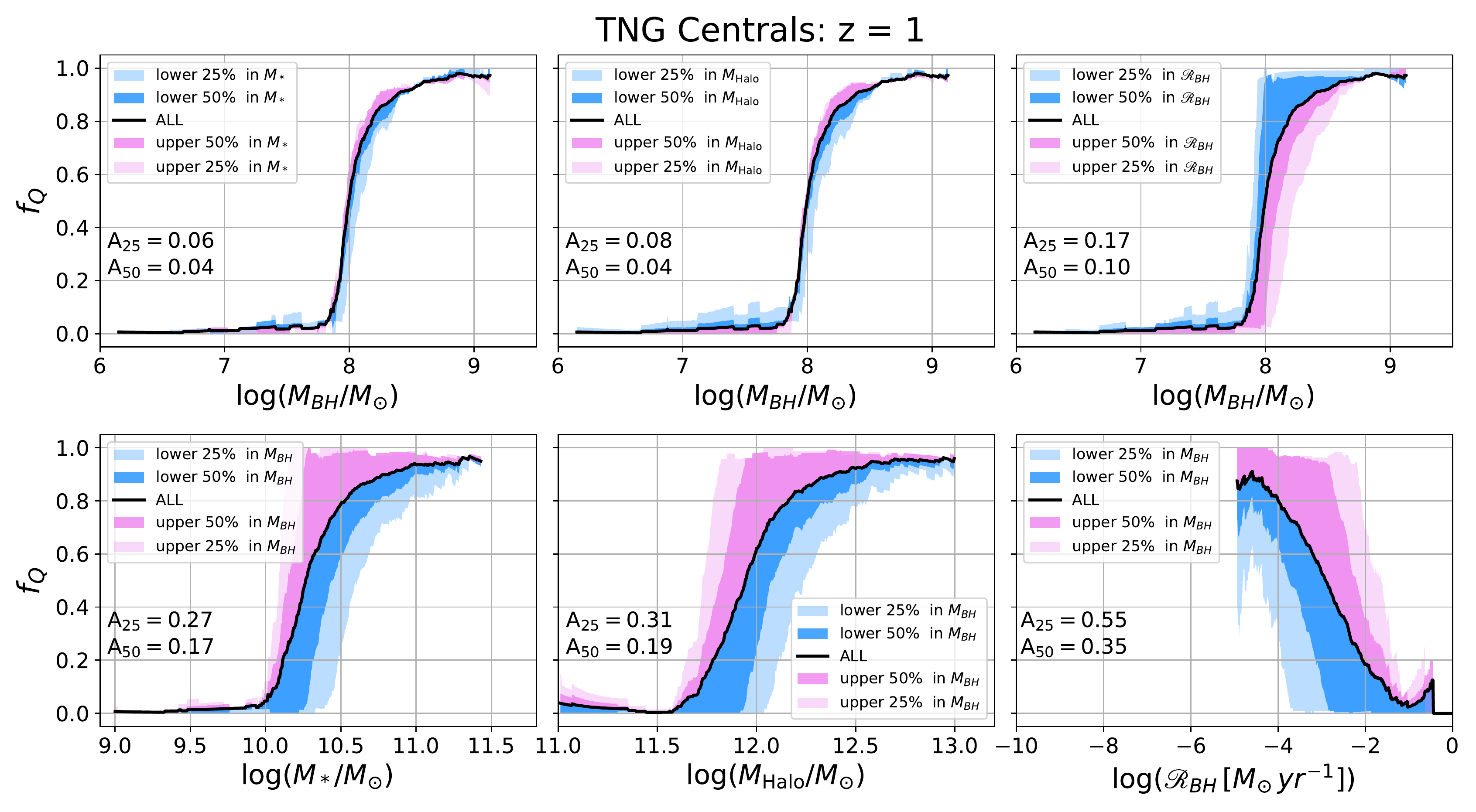}
\includegraphics[width=0.95\textwidth]{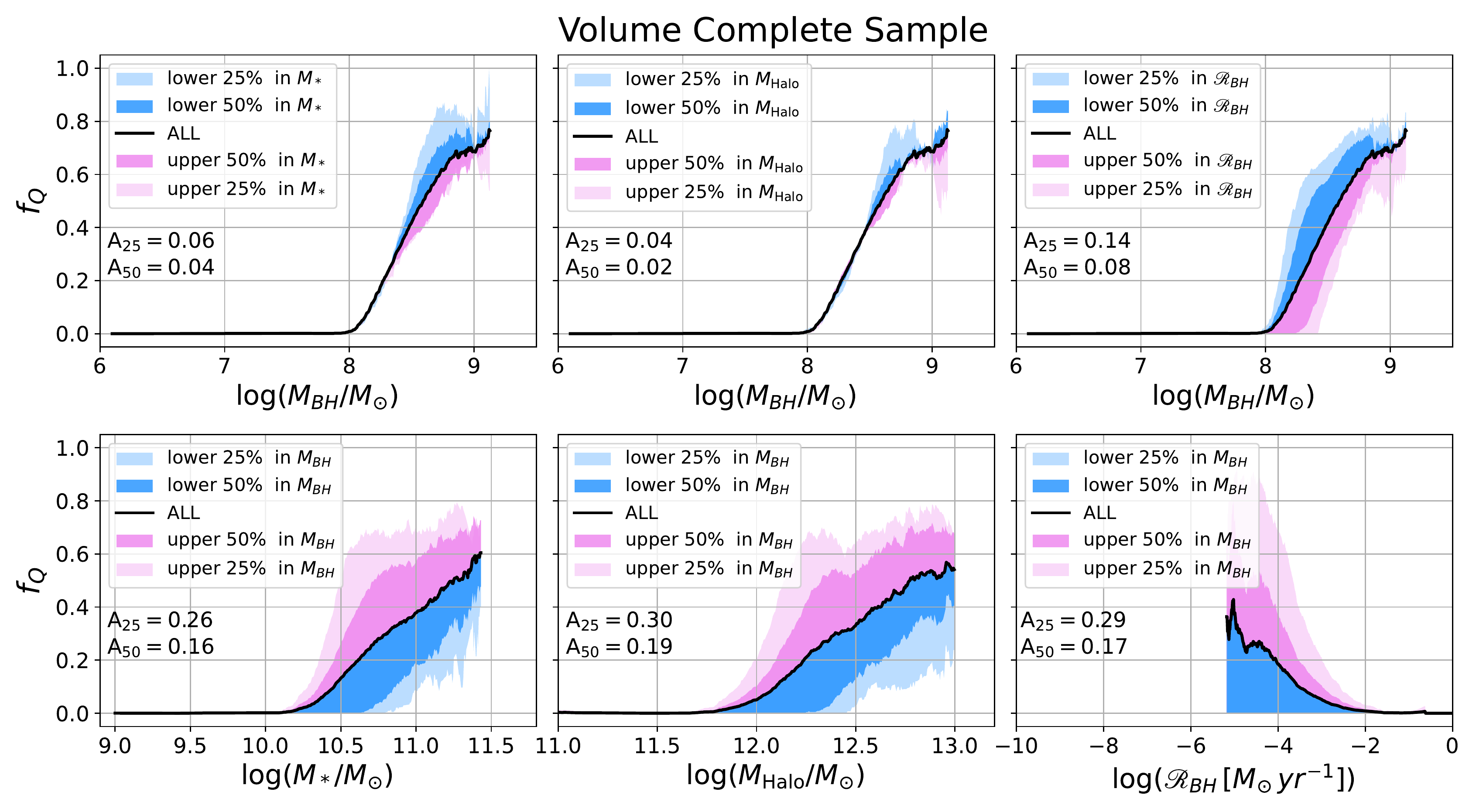}
\caption{{\it Top Panels:} The quenched fraction relationship with black hole mass (top row) and stellar mass, halo mass, and black hole accretion rate (from left-to-right along the bottom row) for the IllustrisTNG simulation at z = 1. Each quenched fraction relationship is split into percentiles based on a third variable (as indicated on the legends). As in the Random Forest analysis, we select a balanced sample (50\% star forming; 50\% quenched) for this test. The tightness of each of the quenched fraction relationships is quantified by the area subtended by the upper and lower 50th (and 25th) percentiles in the third variable (which is displayed on each panel). For example, in the top left panel, the quenched fraction - black hole mass relation is split into ranges based on stellar mass, while the lower left panel inverts this. It is clear that the quenched fraction - black hole mass relation is by far the tightest relationship, confirming the results of the Random Forest classification analysis at this epoch. {\it Bottom Panels: } Same as above, but for a volume complete sample (with a majority of star forming systems). Again, it is clear that the quenched fraction relationship with black hole mass is by far the tightest of the set. }
\end{figure*}

\begin{figure*}
\includegraphics[width=0.95\textwidth]{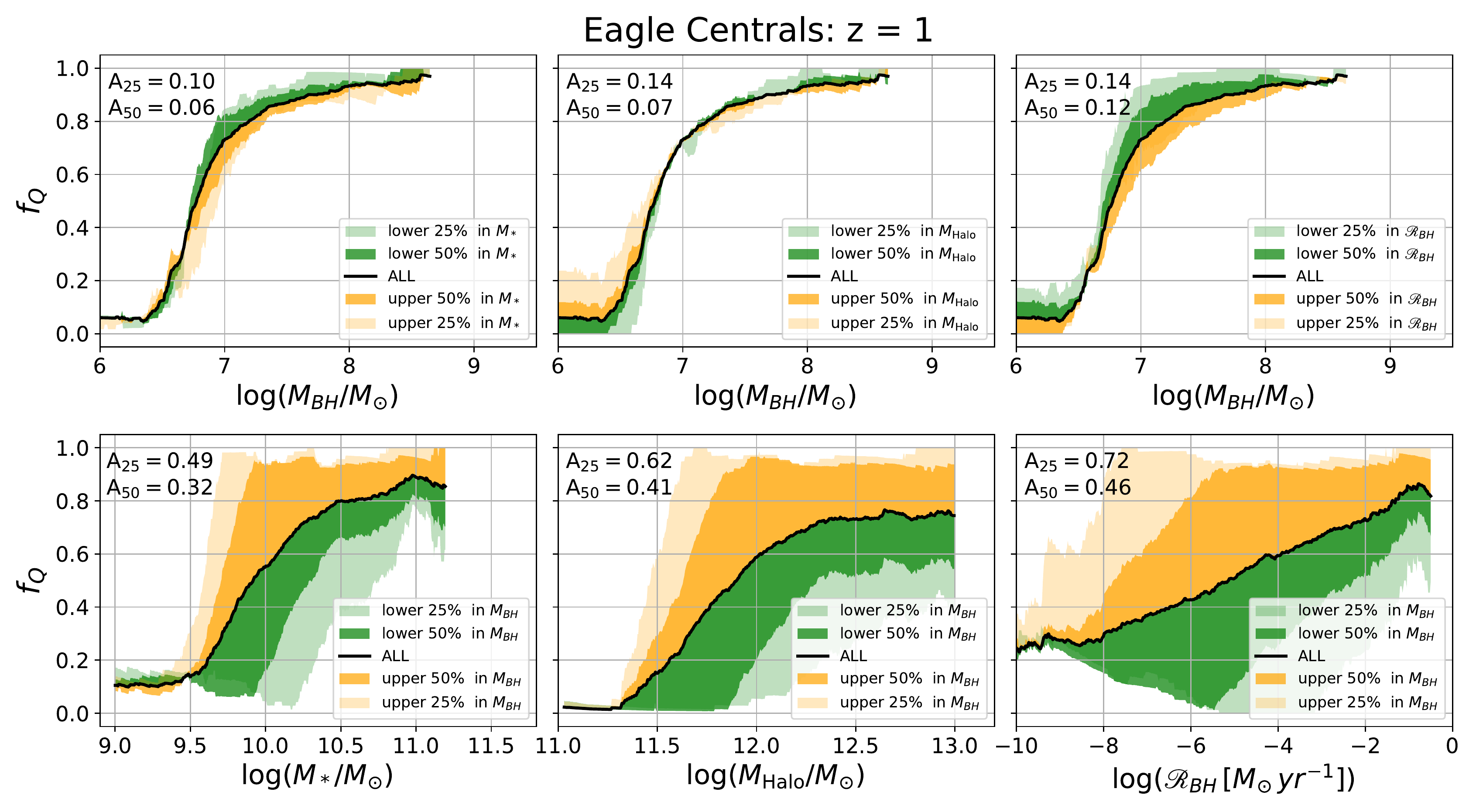}
\includegraphics[width=0.95\textwidth]{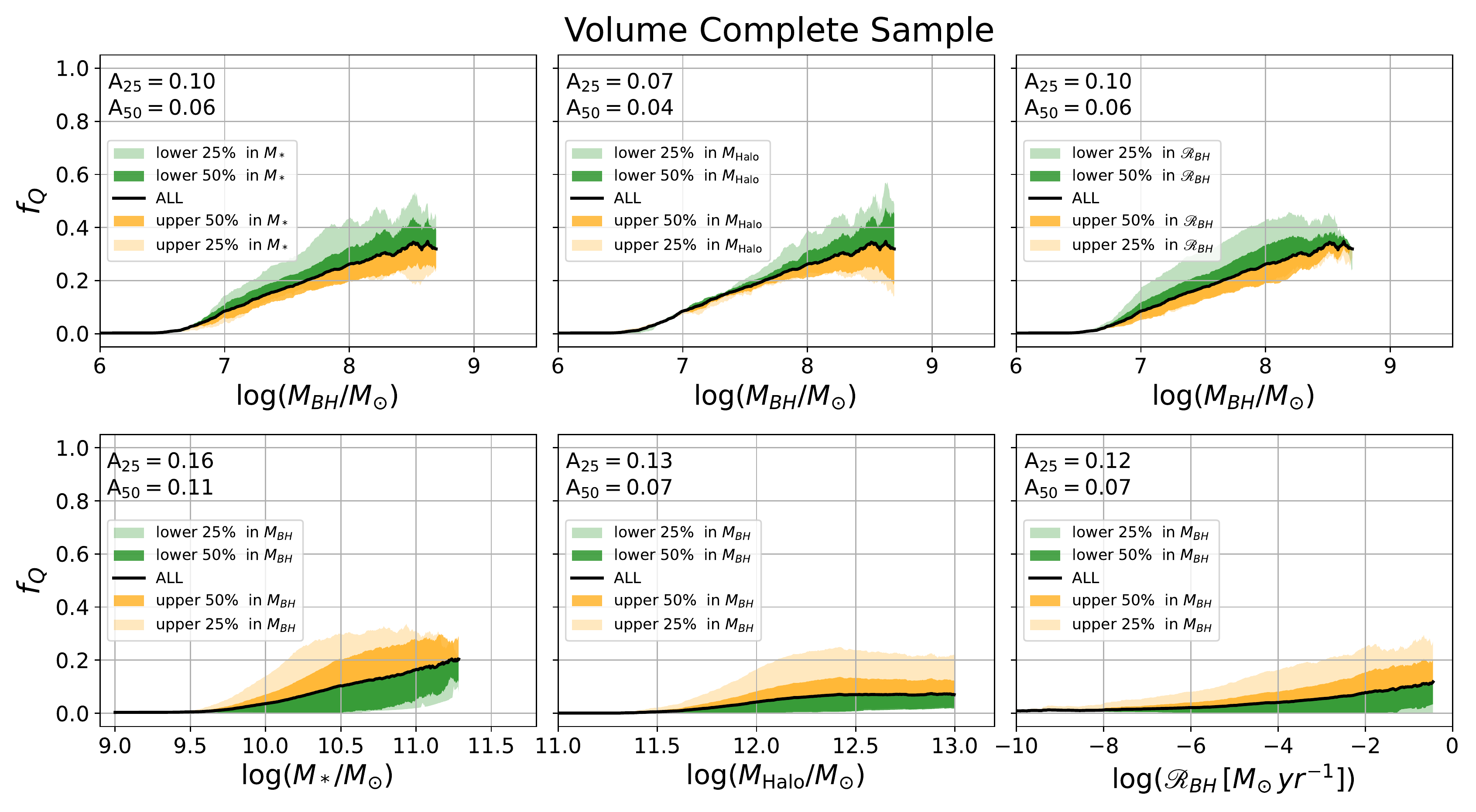}
\caption{The same in structure as Fig. 5, but here showing results for the Eagle simulation at z = 1. The top panels show the results for a balanced sample, and the bottom panels show the results for the volume complete sample. In all cases, the $f_Q - M_{BH}$ relationship is the tightest relation (as quantified by the area statistics displayed on each panel). The only slight exception to this is for $M_{BH}$ vs $\mathscr{R}_{BH}$ in the volume complete sample, where the tightness of each relation is similar. Nonetheless, the increase in quenched fraction is significantly larger as a function of black hole mass than accretion rate. Hence, black hole mass is still a better predictor of quiescence than accretion rate, as seen clearly in the balanced sample (top panel). Moreover, increasing accretion rate at a fixed black hole mass actually {\it lowers} the fraction of quenched galaxies. This is expected for a co-fueling scenario but is the opposite of what is expected for `catching' AGN quenching in action.}
\end{figure*}

\begin{figure*}

\includegraphics[width=0.95\textwidth]{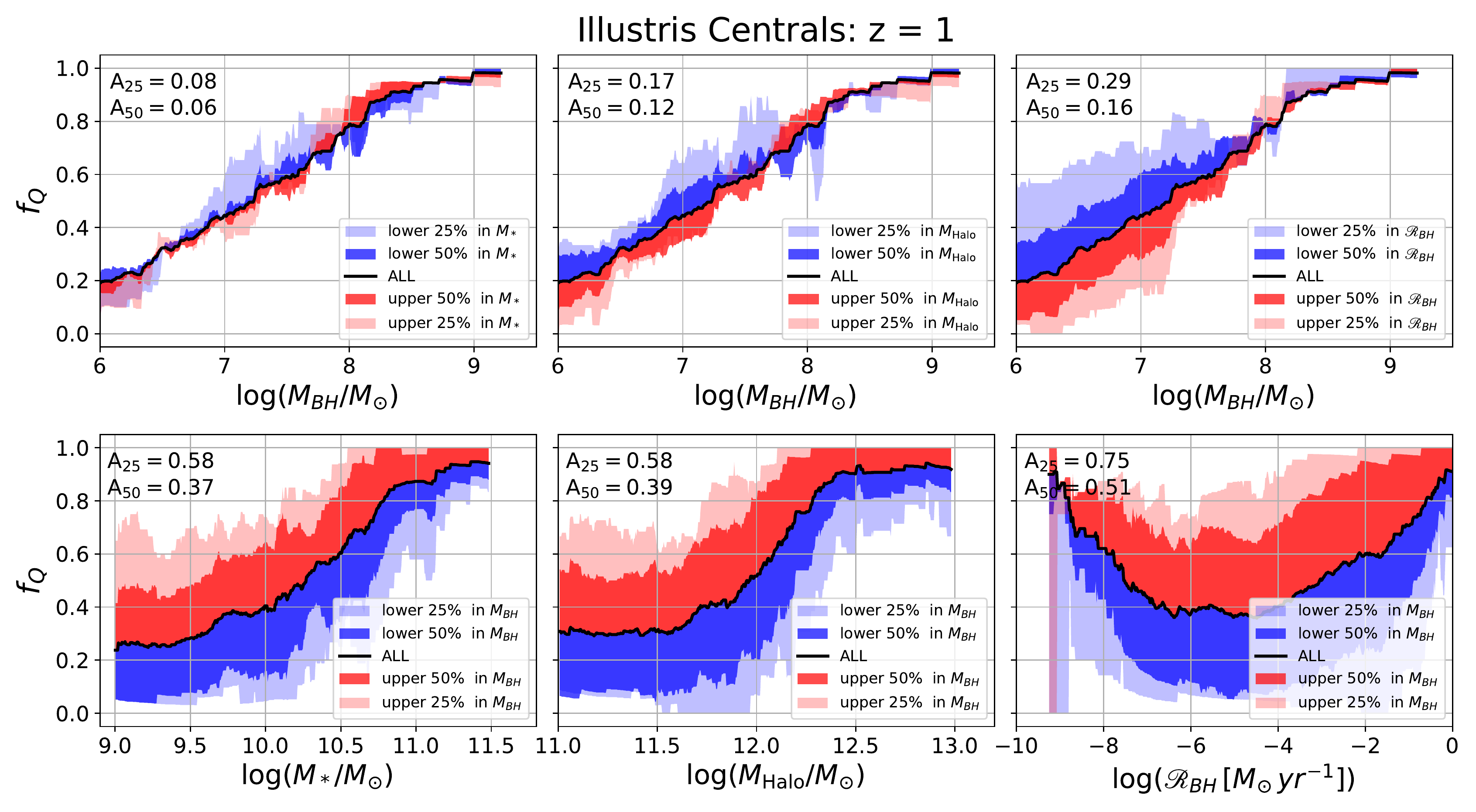}
\includegraphics[width=0.95\textwidth]{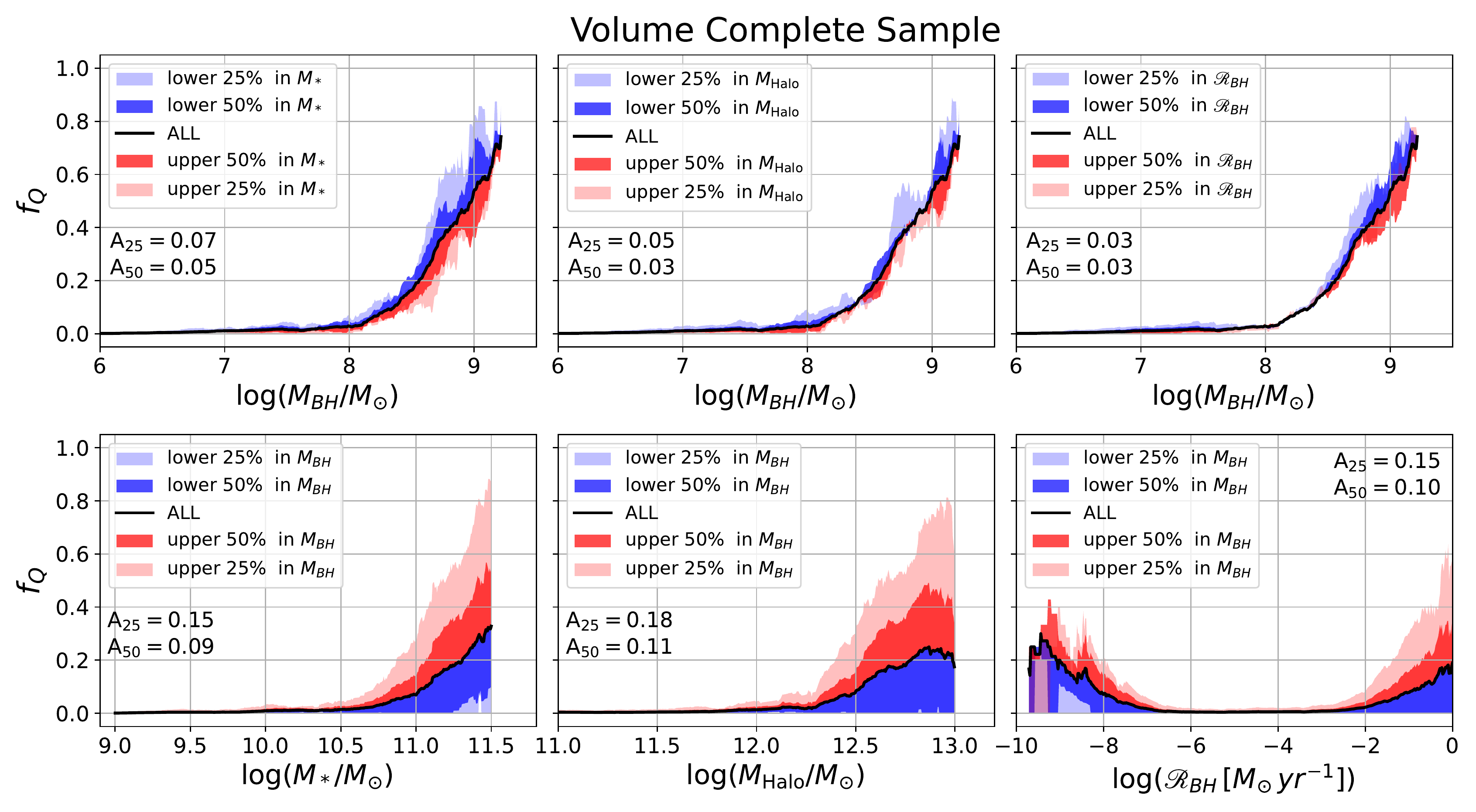}
\caption{The same in structure as Fig. 5, but here showing results for the Illustris simulation at z = 1. The top panels show the results for a balanced sample, and the bottom panels show the results for the volume complete sample. In all cases, the $f_Q - M_{BH}$ relationship is the tightest relation (as quantified by the area statistics displayed on each panel).}
\end{figure*}


\begin{figure*}
\begin{centering}
\includegraphics[width=1\textwidth]{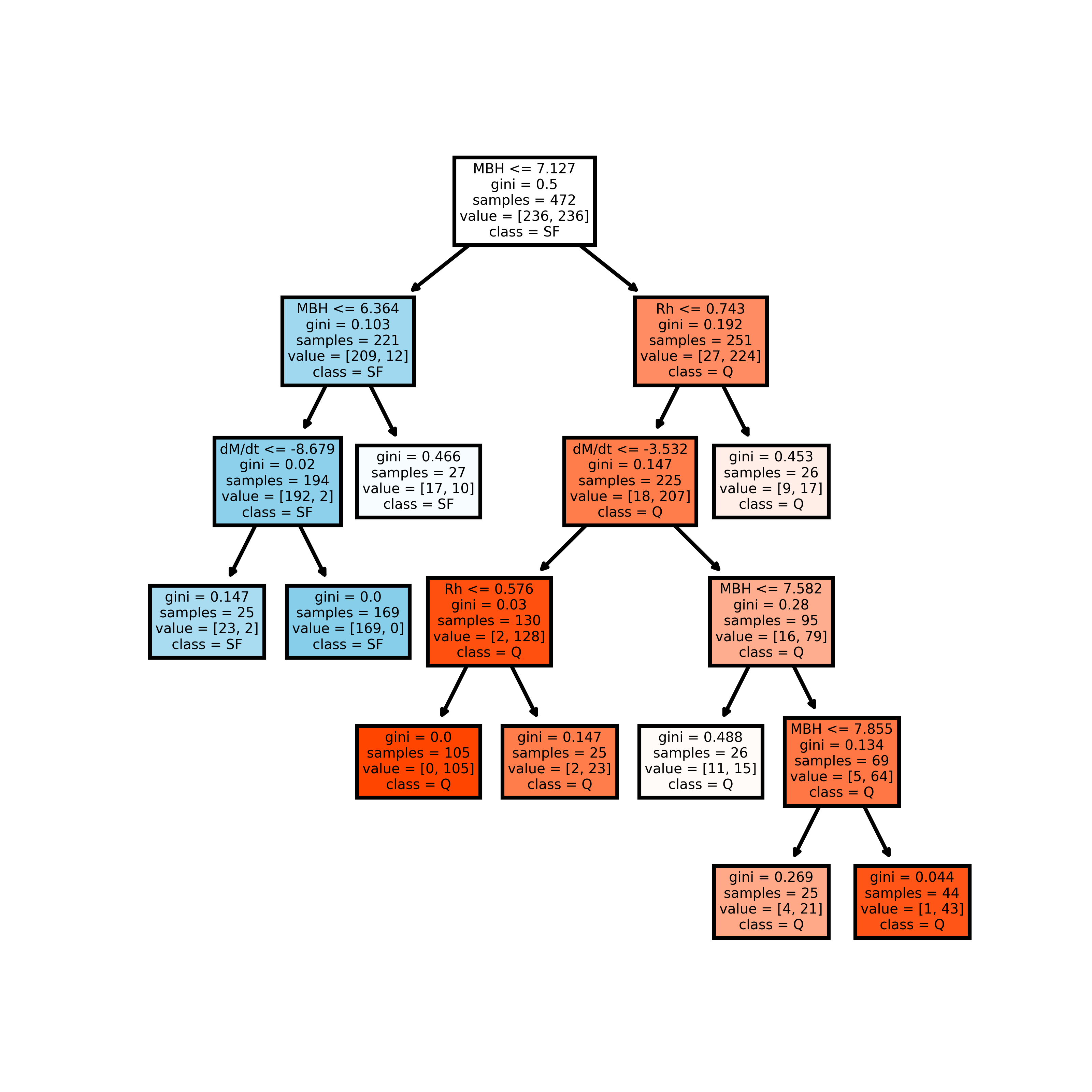}
\caption{An example of one randomly selected decision tree (out of 250) for training in Eagle at z = 1. Here, as in the main body of this paper, we restrict to a balanced sample of quenched and star forming systems. The parameter and threshold chosen by the classifier at each of the decision forks is presented in each box, along with the Gini impurity and sample size (this starts at Gini = 0.5 by design for a balanced sample). The color of boxes indicate whether the descendent sample is majority quenched (red) or star forming (blue), with the depth of the hue indicating the purity of the sample. Colors shift from white to deep blue (or red) down each branch of the decision tree, as the classifier solves the problem. Black hole mass is selected by the classifier as the most effective first cut on the data, engendering the most dramatic reduction in Gini impurity for the largest number of galaxies. The overall success of this parameter emerges from the stability of this choice to sample variation across the entire Random Forest.}
\end{centering}
\end{figure*}

\begin{figure*}
\begin{centering}
\includegraphics[width=1\textwidth]{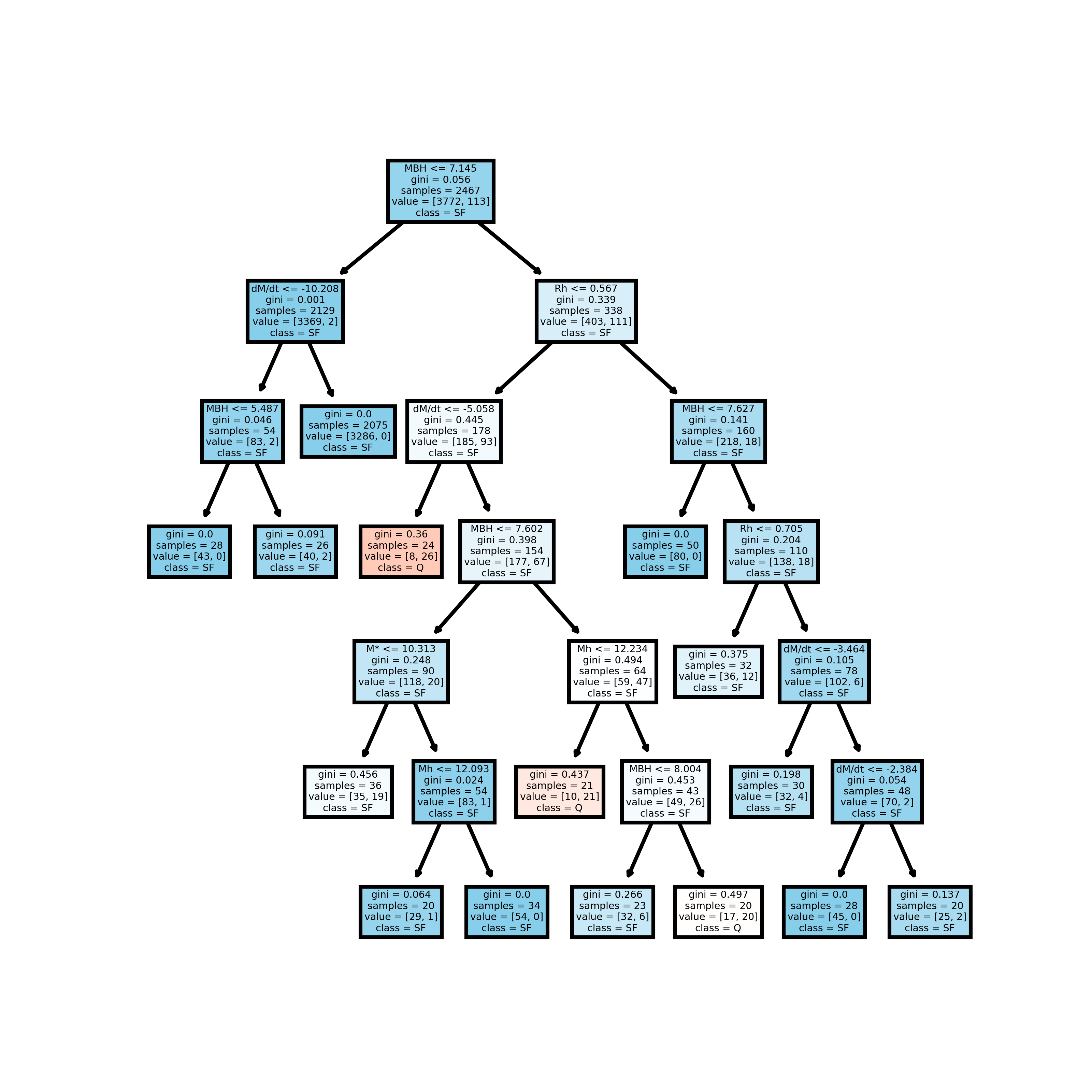}
\caption{An alternative example of a randomly selected decision tree (out of 250) for training in Eagle at z = 1. Here we analyze the raw sample of quenched and star forming systems (which is majority star forming, hence the increased blue shading of this tree compared to Fig. 8). Consequently, the classification problem is more complex and the results require more care to interpret. Nonetheless, black hole mass is still selected as the best parameter to make the first cut on the data, yielding a high performance of black hole mass in this case as well.  }
\end{centering}
\end{figure*}


\begin{figure*}
\begin{centering}
\includegraphics[width=0.32\textwidth]{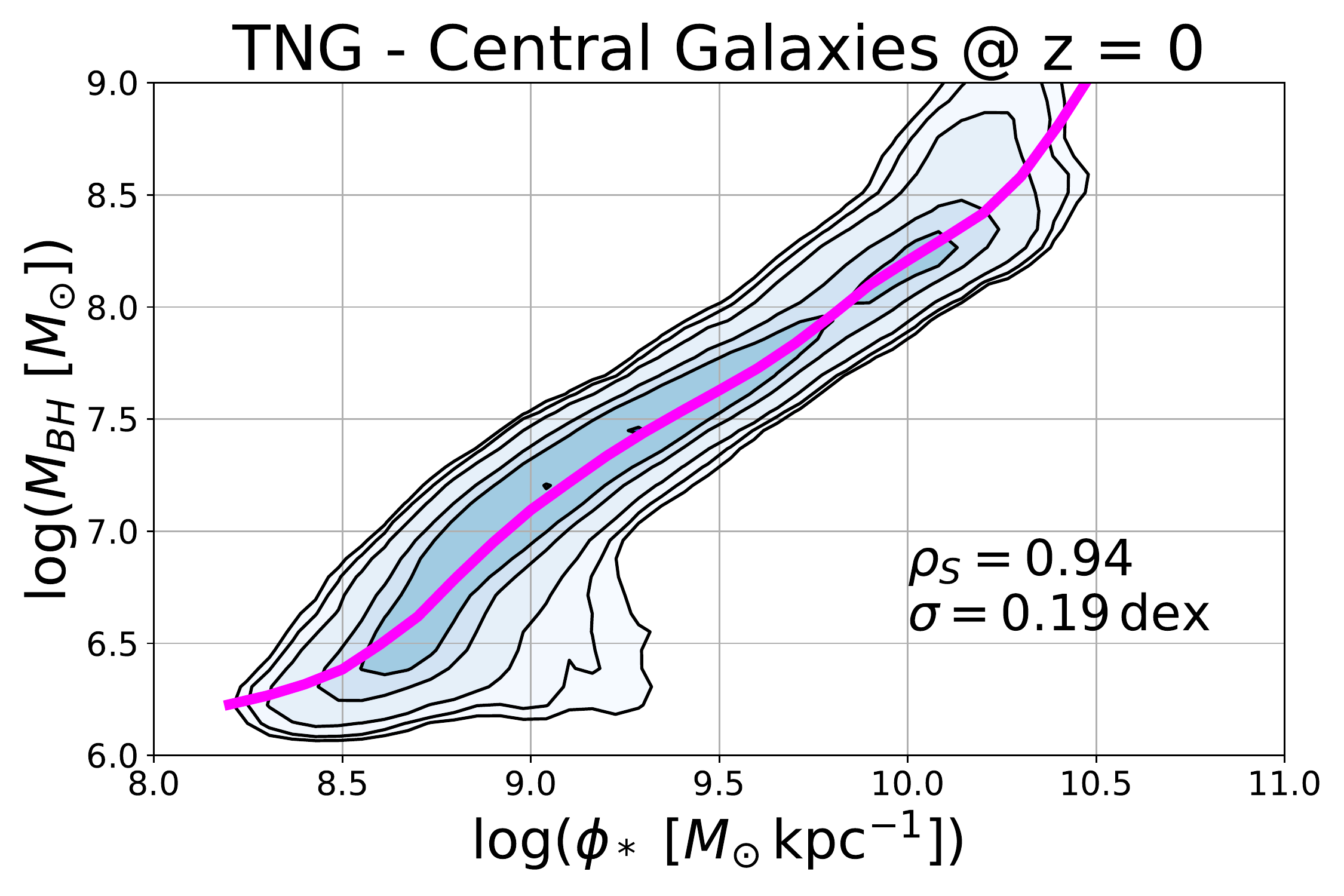}
\includegraphics[width=0.32\textwidth]{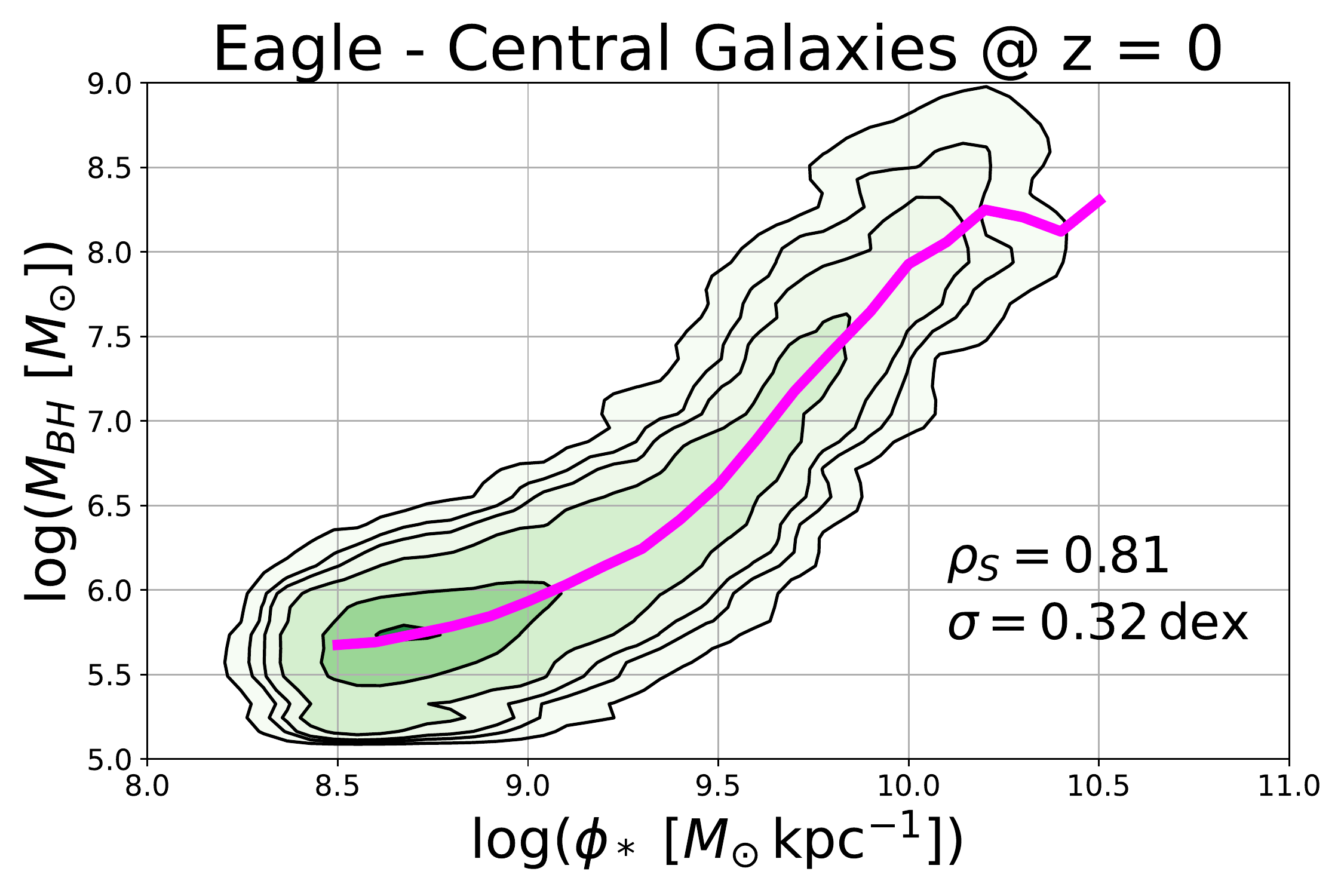}
\includegraphics[width=0.32\textwidth]{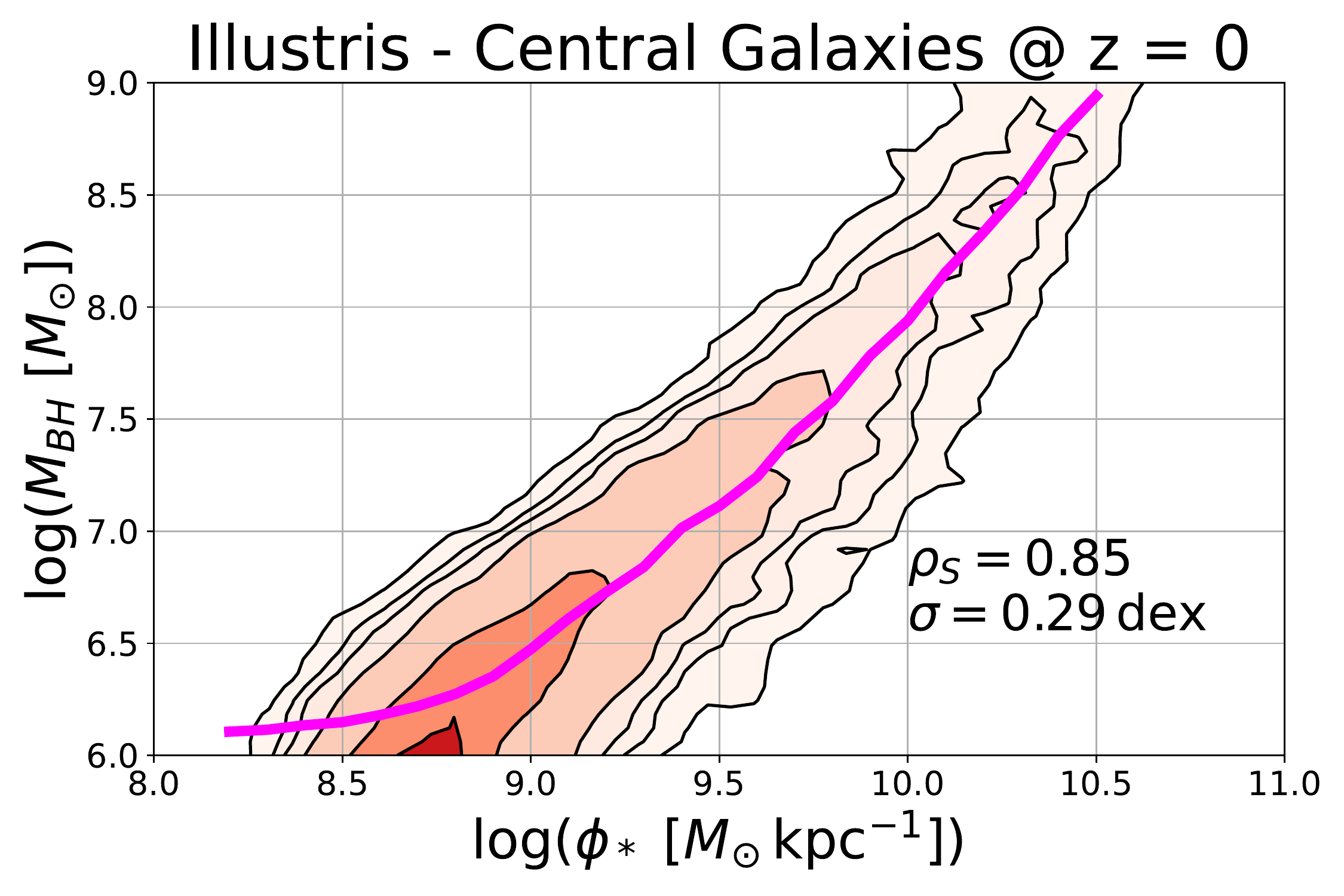}
\includegraphics[width=0.32\textwidth]{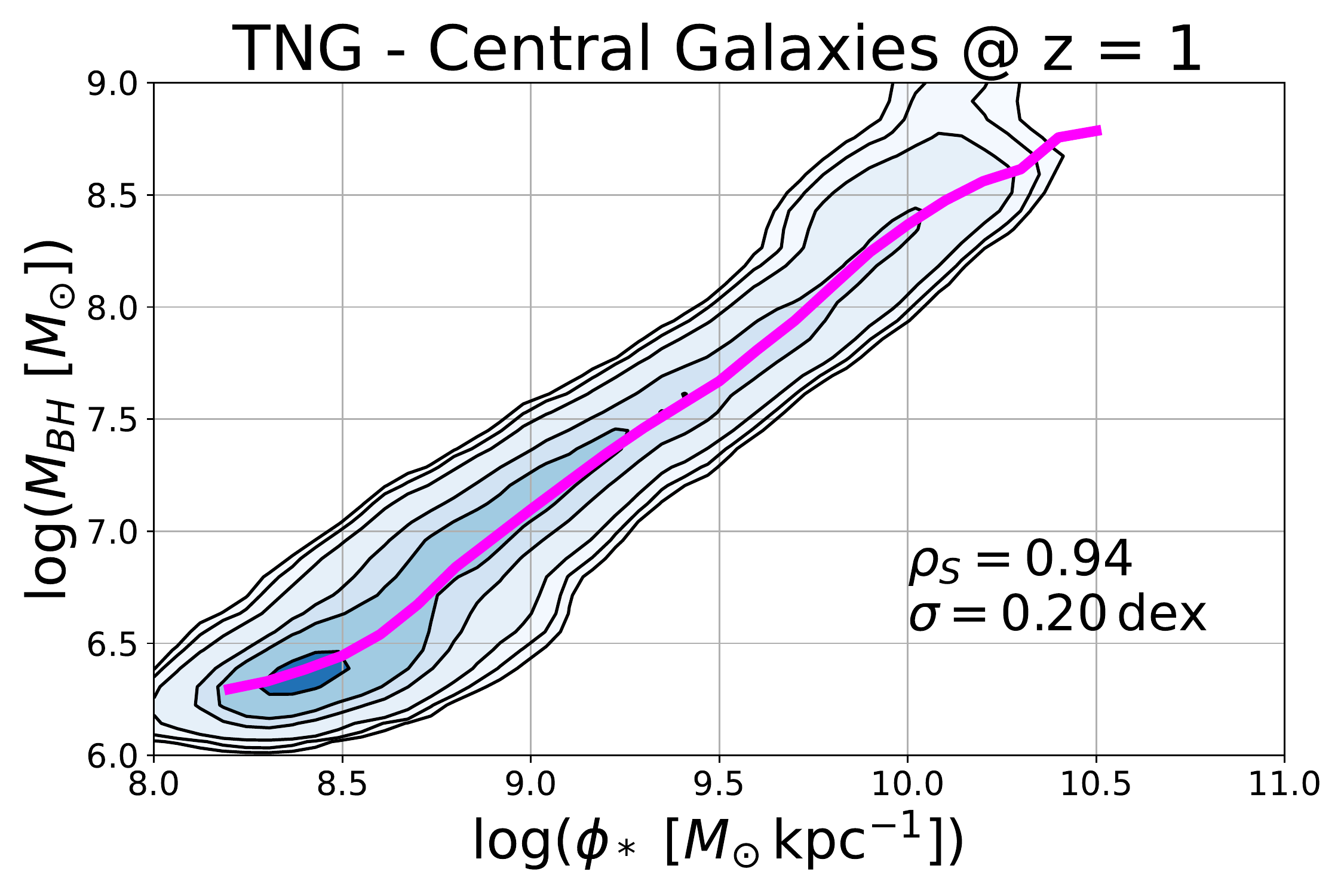}
\includegraphics[width=0.32\textwidth]{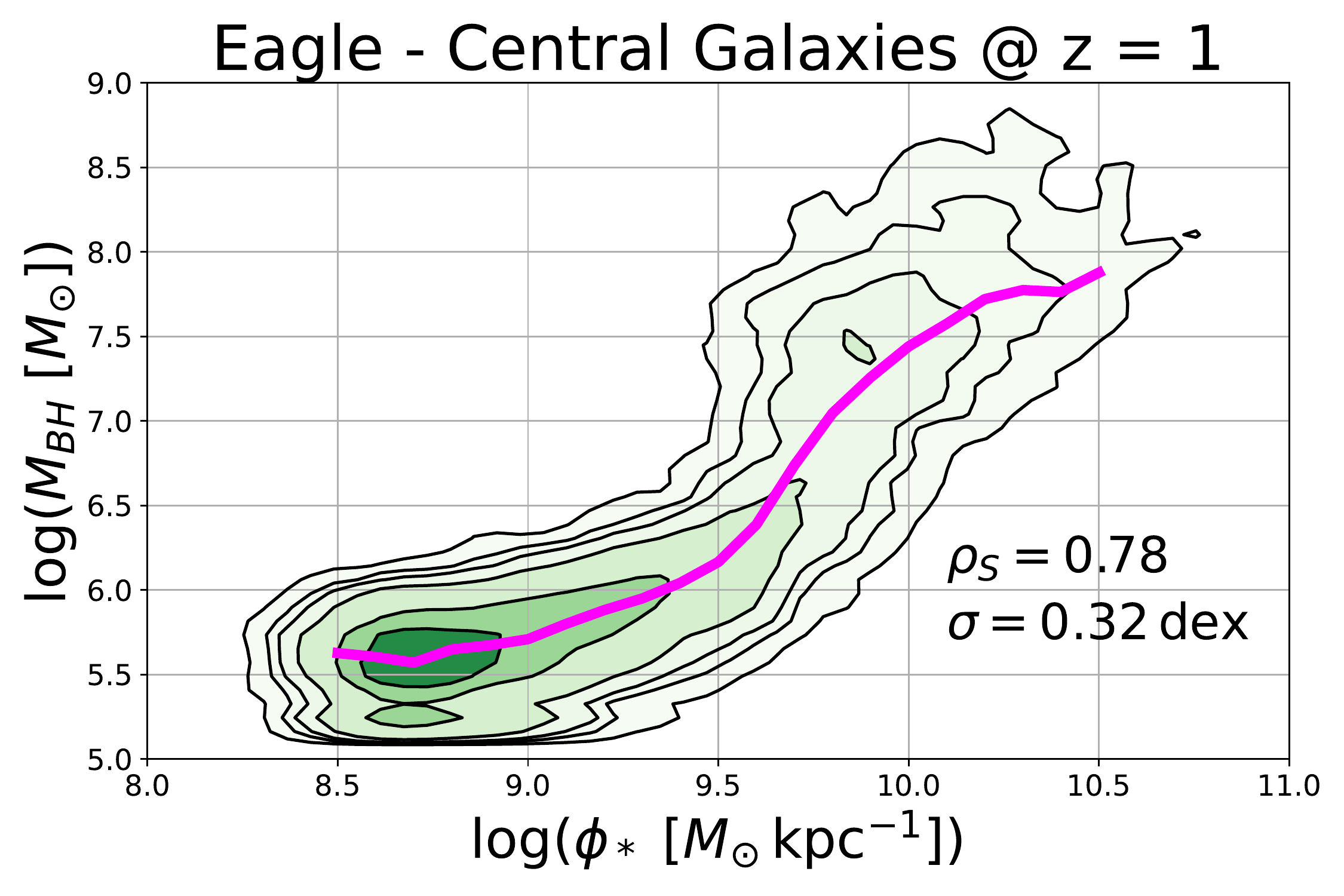}
\includegraphics[width=0.32\textwidth]{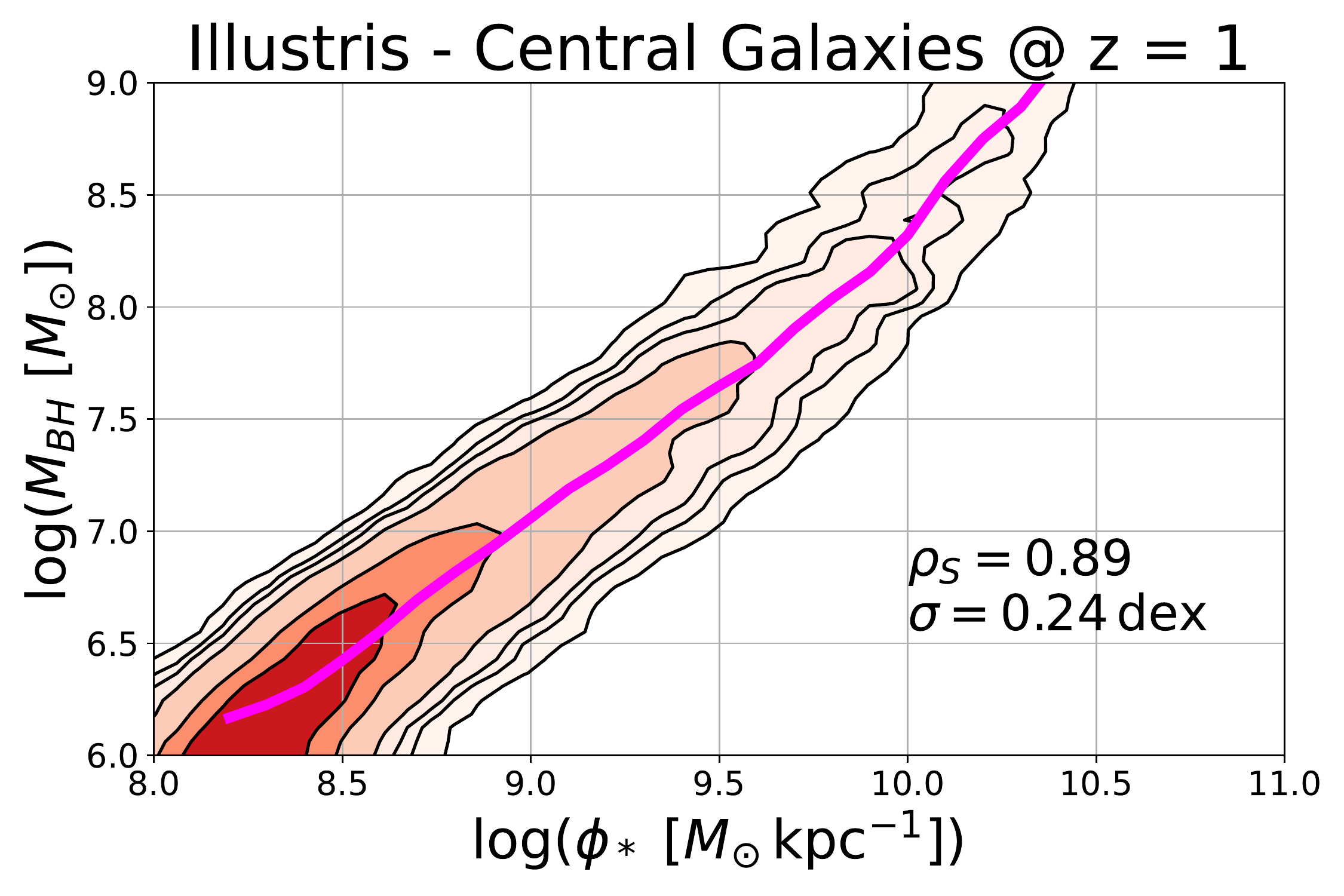}
\includegraphics[width=0.32\textwidth]{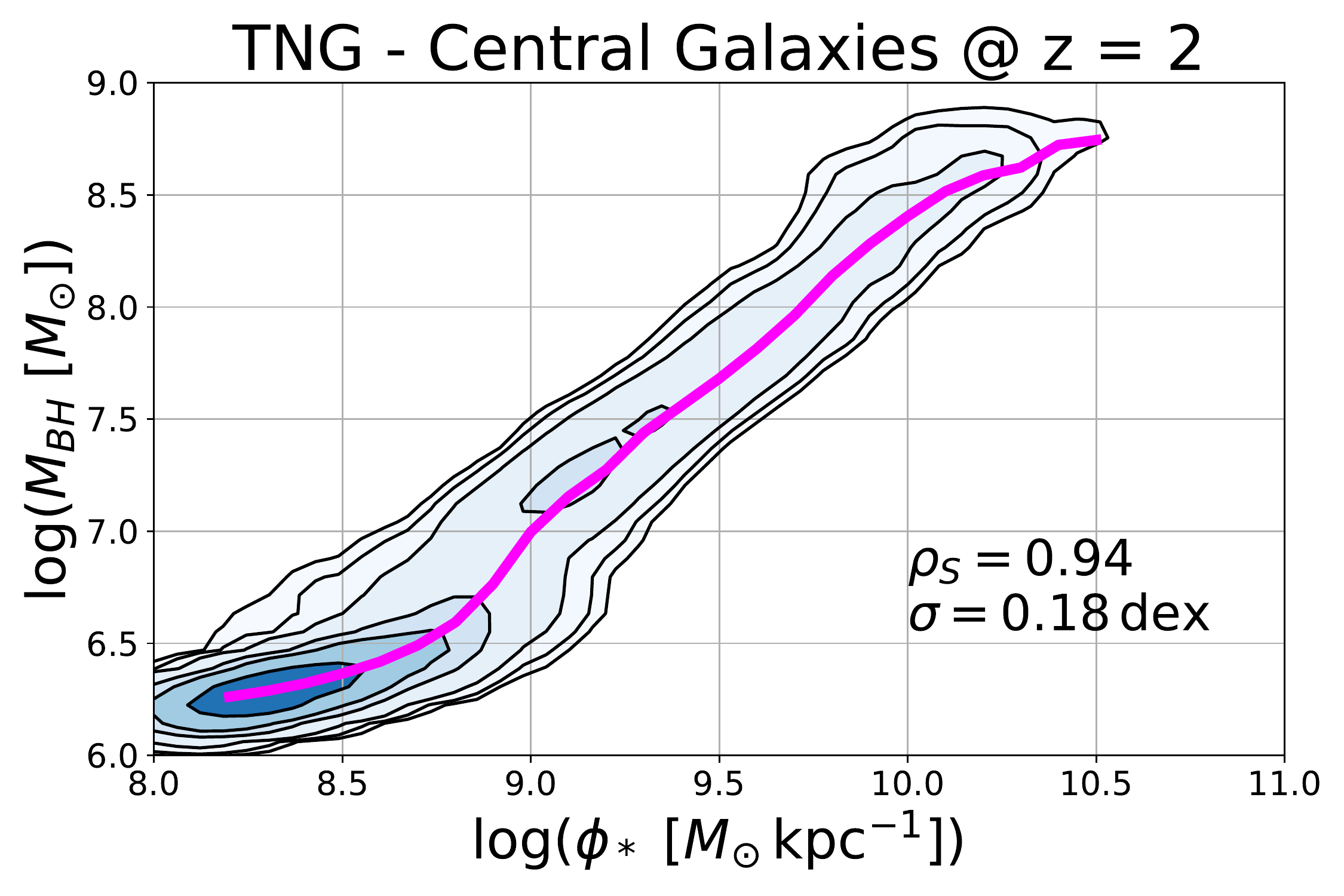}
\includegraphics[width=0.32\textwidth]{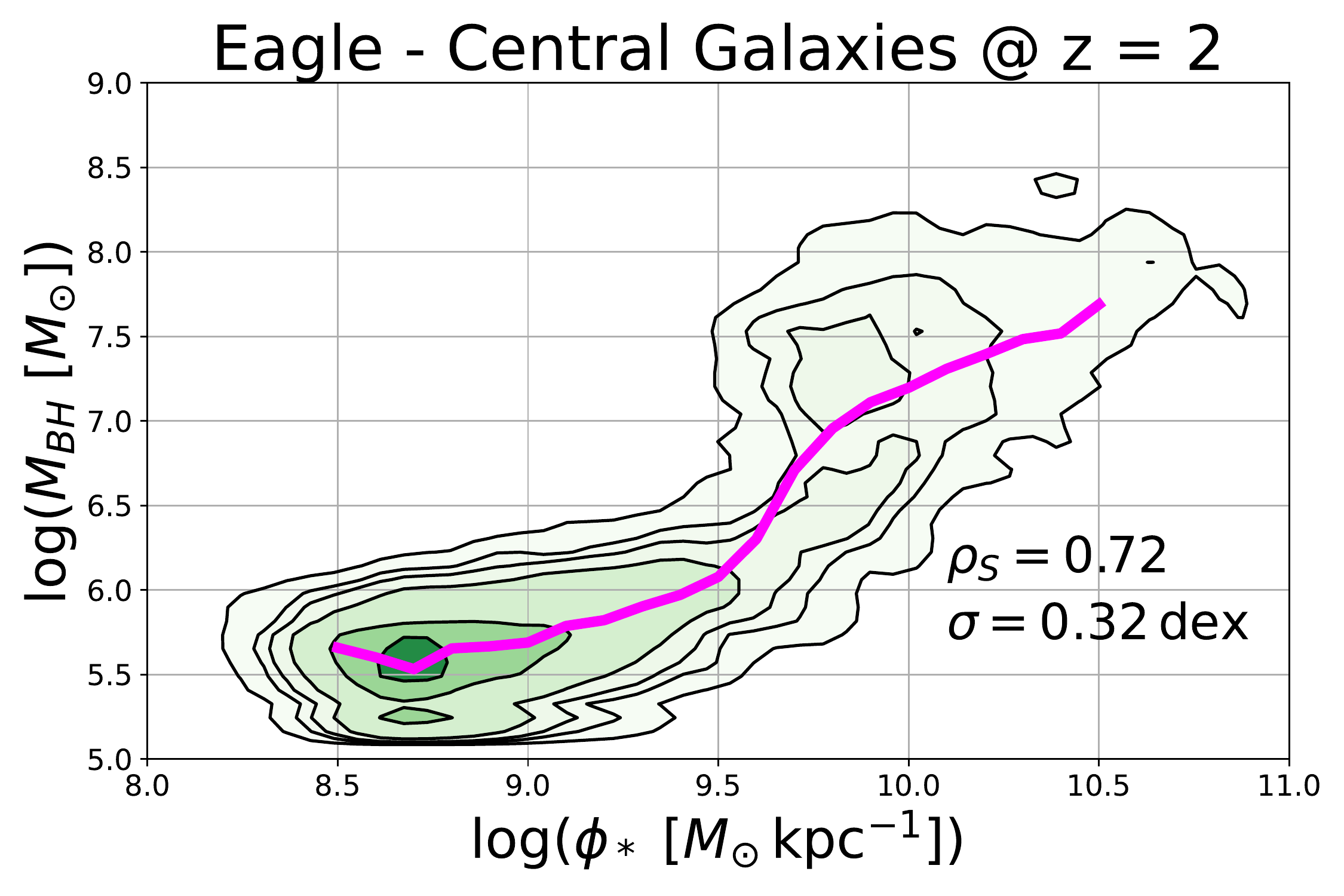}
\includegraphics[width=0.32\textwidth]{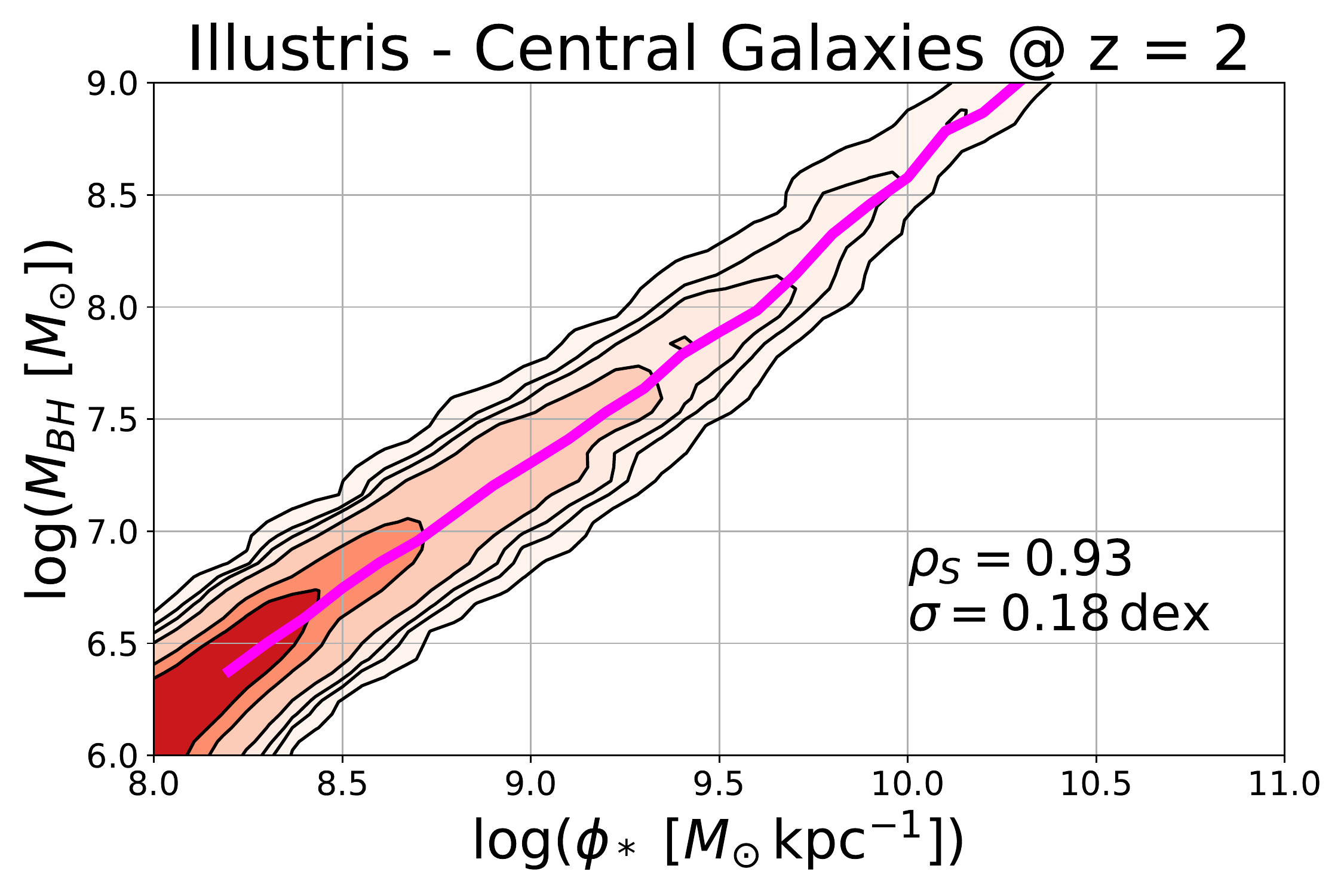}
\caption{Predicted relationships between black hole mass and the stellar potential for TNG (left panels), Eagle (middle panels) and Illustris (right panels). Results are shown for z = 0 (top row), z = 1 (middle row), and z = 2 (bottom row). The median relationship is indicated by a solid magenta line, and the spread of the data is indicated by linearly spaced contours. The Spearman correlation statistic ($\rho_S$) and the dispersion about the median relation ($\sigma$) are displayed on each panel. There is predicted to be a strong and reasonably tight relationship between black hole mass and the stellar potential by all three simulations at all epochs. Hence, $\phi_*$ is expected to act as a reasonable proxy for $M_{BH}$ in observational data as well.}
\end{centering}
\end{figure*}


\begin{figure*}

\includegraphics[width=0.9\textwidth]{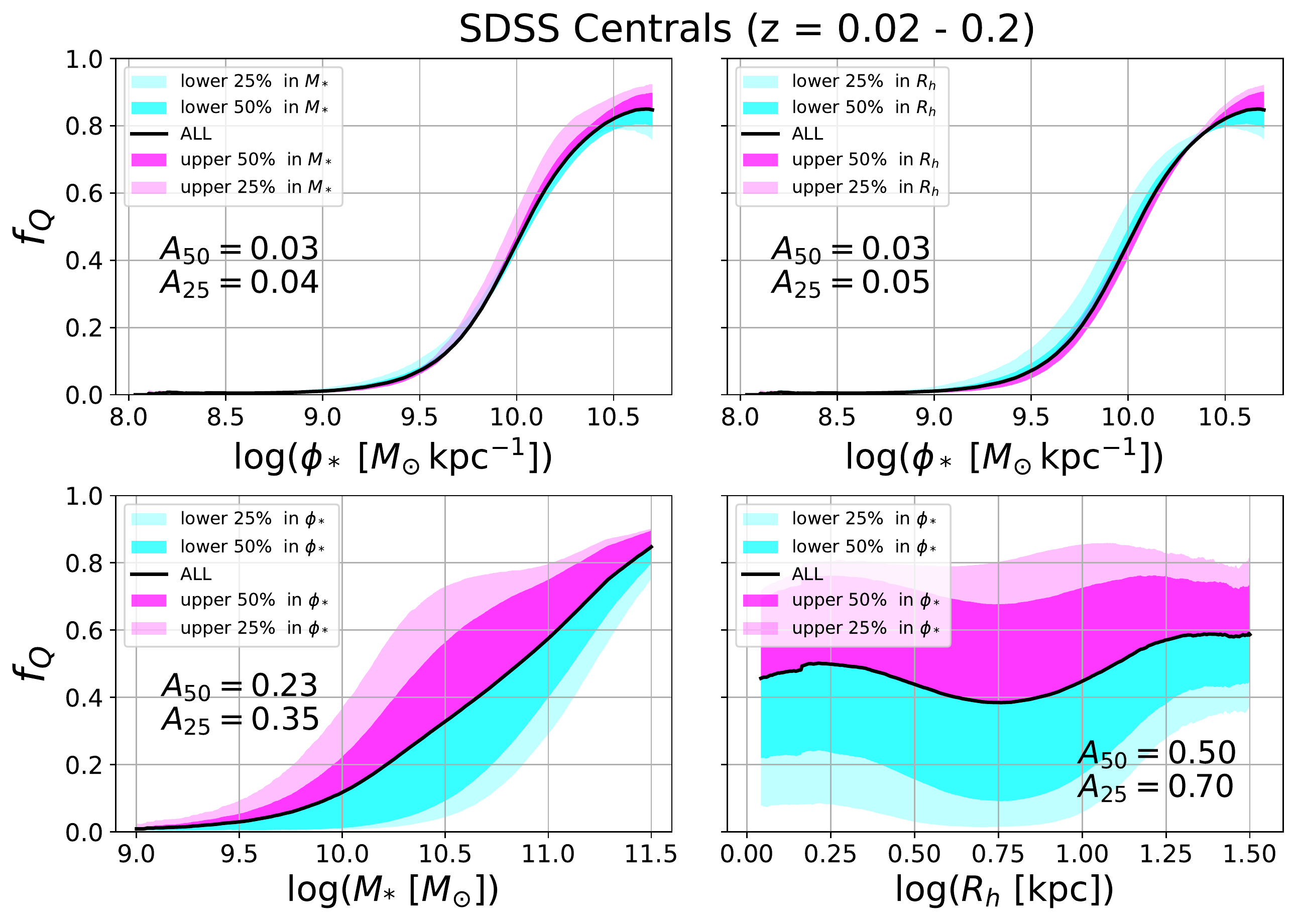}
\caption{The quenched fraction relationship with $\phi_*$ (top panels), $M_*$ (bottom-left panel), and $R_h$ (bottom-right panel) in the SDSS. As in Fig. 5, each quenched fraction relationship is displayed for percentile ranges of a third parameter, in order to assess the impact on quenching of varying that parameter at fixed values of each of the others. The tightness of each relationship is quantified by the area statistics displayed on each panel. It is clear that $\phi_*$ engenders the tightest and steepest of the quenched fraction relationships, exactly as predicted in models utilizing AGN feedback to quench galaxies.}. 
\end{figure*}

\begin{figure*}
\includegraphics[width=0.9\textwidth]{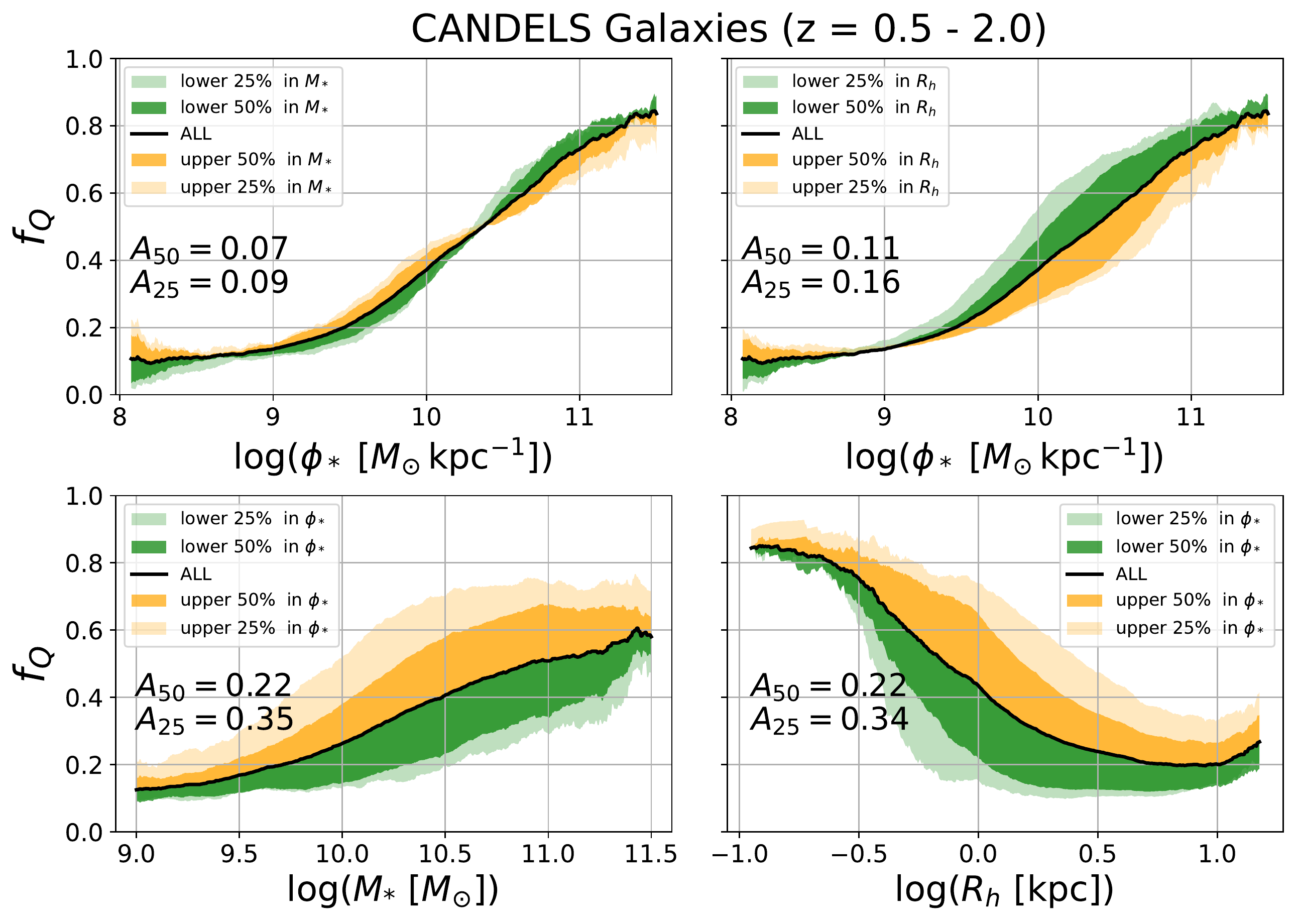}
\caption{Identical in structure to Fig. 11, but here showing results from CANDELS. As at low redshifts, $\phi_*$ exhibits the tightest quenched fraction relationship, as quantified by the area statistics displayed on each panel. Note also that $\phi_*$ exhibits the steepest positive relationship with quiescence.}
\end{figure*}

\end{document}